\documentclass[twocolumn,aps,prd,amsmath,amssymb,floatfix,amsmath,superscriptaddress]{revtex4-2}

\usepackage{graphicx}% Include figure files
\usepackage{dcolumn}% Align table columns on decimal point
\usepackage{bm}% bold math
\usepackage{braket}
\usepackage{color,epsfig}
\usepackage{bm}
\usepackage{slashed}
\usepackage{hyperref}
\usepackage{subfigure}
\usepackage{feynmf}
\usepackage{calrsfs}
\usepackage[mathscr]{euscript}
\usepackage{amssymb}
\usepackage{booktabs}
\usepackage{siunitx}
\usepackage{xcolor}

\newcommand{\bea}{\begin{eqnarray}}
\newcommand{\eea}{\end{eqnarray}}
\newcommand{\be}{\begin{equation}}
\newcommand{\ee}{\end{equation}}

\begin{document}

%\preprint{}

\title{VSL-Gravity in light of PSR B1913+16 Full Data Set: \\
Upper limits on graviton mass and its theoretical consequences}

\author{Alexander Bonilla}
\email{alexander.bonillarivera@nanograv.or}
\affiliation{Observat\'orio Nacional, Rua General Jos\'e Cristino 77, S\~ao Crist\'ov\~ao, 20921-400 Rio de Janeiro, RJ, Brazil}
\affiliation{Instituto de F\'{i}sica, Universidade Federal Fluminense, 24210-346 Niter\'{o}i, RJ, Brazil}
 
\author{Alessandro Santoni}
\email{asantoni@uc.cl}
\affiliation{Institut f\"ur Theoretische Physik and Atominstitut,
 Technische Universit\"at Wien,
 Wiedner Hauptstrasse 8--10,
 A-1040 Vienna, Austria}
\affiliation{Instituto de Física, Pontificia Universidad de Católica de Chile, \\
 Avenida Vicuña Mackenna 4860, Santiago, Chile }

\author{Rafael C. Nunes}
\email{rafadcnunes@gmail.com}
\affiliation{Instituto de F\'{i}sica, Universidade Federal do Rio Grande do Sul, 91501-970 Porto Alegre RS, Brazil}
\affiliation{Divis\~ao de Astrof\'isica, Instituto Nacional de Pesquisas Espaciais, Avenida dos Astronautas 1758, S\~ao Jos\'e dos Campos, 12227-010, SP, Brazil}

\author{Jackson Levi Said}
\email{jackson.said@um.edu.mt}
\affiliation{Institute of Space Sciences and Astronomy, University of Malta, Msida, Malta}
\affiliation{Department of Physics, University of Malta, Msida, Malta}

\date{\today}% It is always \today, today, but any date may be explicitly specified

\begin{abstract}
Very Special Linear Gravity (VSL-Gravity) is an alternative model of linearized gravity that incorporates massive gravitons while retaining only two physical degrees of freedom thanks to gauge invariance. Recently, the gravitational period-decay dynamics of the model has been determined using effective field theory techniques. In this study, we conduct a comprehensive Bayesian analysis of the PSR B1913+16 binary pulsar dataset to test the predictions of VSL-Gravity. Our results place a 95\% confidence level upper bound on the graviton mass at $m_g \lesssim  10^{-19} \, \text{eV}/c^2$. Additionally, we observe a significant discrepancy in the predicted mass of one of the binary's companion stars. Lastly, we discuss the broader implications of a non-zero graviton mass, from astrophysical consequences to potential cosmological effects.
\end{abstract}

\maketitle

%\tableofcontents
\section{Introduction}
Since the pioneering work of Fierz and Pauli \cite{Fierz:1939ix} in 1939, many have struggled to describe massive gravitational models while avoiding dangerous pathologies. Massive gravity theories are often plagued by additional ghost-like degrees of freedom (DOFs), which compromise unitarity, as well as discontinuities, such as the van Dam-Veltman-Zakharov (vDVZ) discontinuity \cite{vanDam:1970vg,Zakharov:1970cc}, because of which the massless limit does not recover General Relativity (GR). Nevertheless, interest in these alternatives to GR has grown in recent years due to their potential to address some of the major open questions in modern physics, including dark energy \cite{hinterbichler2012theoretical} and cosmological tensions \cite{de2021minimal}. Moreover, solutions to the aforementioned pathologies, such as the Vainshtein mechanism \cite{Vainshtein:1972sx}, have been explored and implemented in various models, for instance, the DGP model \cite{Dvali:2000hr,Dvali:2000rv} and dRGT gravity \cite{deRham:2010kj}. 
%Massive gravity framework has been tested and explored in the astrophysical and cosmological contexts in \cite{} \textbf{...}. 

Recently, Alfaro and Santoni proposed a new massive model for linearized gravity, which they called Very Special Linear Gravity (VSL-Gravity or VSLG) \cite{grav3}. The latter is based on a peculiar Lorentz-violating realization developed in 2006 by Cohen and Glashow, denoted Very Special Relativity (VSR) \cite{vsr1}. This was originally introduced as an alternative mechanism for neutrino masses \cite{vsr2}, but was soon implemented in other contexts and was found to generally enable the description of massive gauge DOFs \cite{vsrqed,Alfaro:2013uva}. The unique feature of VSR as a Lorentz-violating approach is the absence of additional invariant tensors in the structure of the new spacetime symmetry group. This indicates that VSR effects cannot be parametrized through spurionic fields and inhevitably lead to non-localities \cite{Santoni:2024coa}. Nevertheless, VSR still allows for a null preferred direction in spacetime, in the sense that there exists a lightlike vector $n^\mu$ that only gets rescaled under the action of the VSR group. Therefore, new VSR-invariant contributions are usually constructed as ratios of scalar products containing $n^\mu$ both in the numerator and in the denominator. Moreover, when working in flat space or in the linearized approximation, this term structure allows us to normalize the vector as $n^\mu = (1, \hat n)$, leaving just the generic spatial direction $\hat n$ to be determined.

Within the massive gravity paradigm, VSL-Gravity exhibits novel features that could be of interest in various areas of gravitational physics and set it apart from other viable candidates. Specifically, it describes gauge-invariant, yet massive, gravitons without introducing the usual “unhealthy” DOFs seen in massive gravity formulations. This advantage, however, comes at the cost of incorporating a preferred null spacetime direction, denoted by the four-vector $n^\mu$, leading to a certain degree of anisotropy \cite{vsr1}. Due to the absence of a complete non-linear extension of VSL-Gravity, the physical interpretation of $n^\mu$ in curved spacetime remains unclear. This anisotropy could either be attributed to a cosmological preferred direction, possibly originating from remnants of a primordial magnetic field permeating the Universe \cite{Giovannini:2002sv}, or to more local \cite{background} and fundamental features of spacetime itself \cite{Das:2018umm,Mann:2020jcu}. 

Since their discovery by Hulse and Taylor in 1974 \cite{1975ApJ...195L..51H}, binary pulsars have played a crucial role in testing the predictions of General Relativity (GR), serving as the first indirect evidence for gravitational waves (GWs), as Taylor \& Weisberg showed us in their seminal 1982 work \cite{1982ApJ...253..908T}. Increasingly precise measurements of orbital decay dynamics have allowed these systems to be used as powerful tools for discriminating between GR and alternative theories or, at the very least, placing stringent constraints on them. Such analyses have already been carried out for various models, including more conventional massive gravity theories \cite{Poddar:2021yjd} and Galileon models \cite{shao2020new}.
%, among others \cite{}. \textbf{...}. 

The aim of this work is to perform a comprehensive statistical and Bayesian analysis of VSL-Gravity predictions regarding the orbital decay of the PSR B1913+16 binary system. Through this analysis, we specifically seek to extract information on the possible values of the graviton mass $m_g$ within the VSL-Gravity framework, while simultaneously placing constraints on this parameter. Further investigation concerning the preferred direction $n^\mu$ in VSL-Gravity is left for future studies.

The outline of this paper is as follows: In Section \ref{model}, we provide a brief introduction and description of the formula used for the period loss rate in a binary system due to VSL-Gravity gravitational radiation. In Section \ref{results}, we describe our methodology for analyzing the data and present our main results. In Section \ref{consequences}, we interpret the results and discuss their implications in both astrophysical and cosmological contexts. Finally, in Section \ref{final}, we conclude with some final remarks and future perspectives.

\section{Gravitational Orbital Decay in VSL-Gravity}
\label{model}

The emission of gravitational waves during the motion of a binary system with masses $m_1$ and $m_2$ inevitably leads to orbital decay, resulting in a decrease in the orbital period $P$ and an increase in the fundamental frequency of motion, $\Omega \equiv \frac{2\pi}{P}$. In a Keplerian system, the rate of period decrease, $\frac{dP}{dt}$, is directly related to the energy emission rate, $\frac{dE}{dt}$, as: \cite{1975ApJ...195L..51H,Weisberg:2010zz,grav3}

\begin{equation}\label{Eq:dP_dE}
    \frac{dP}{dt}= - 6 \pi \frac{ b^{\frac52 }G^{-\frac32} }{ m_1 m_2 \sqrt{m_1+m_2} } \frac{dE}{dt} \, ,
\end{equation}
with $b$ denotes the semi-major axis of the orbit, and $G$ is the Newtonian constant of gravitation. The calculation of the gravitational energy emission rate, $\frac{dE}{dt}$, can be approached through various methods. Notably, in \cite{Santoni:2023uko}, the authors employed effective field theory techniques to determine this quantity within the framework of VSL-Gravity. In standard GR, the formula for the period derivative, $\frac{dP}{dt}$, was first derived and presented in 1963 by Peters and Mathews
\cite{peters1963gravitational} and read
\begin{equation}
    \frac{dP_{GR}}{dt} = -\frac{192 \pi \, T_{\odot}^{\frac53}}{5}  \frac{\tilde m_1 \tilde m_2}{\tilde M ^{\frac13}} \left (\frac{P}{2\pi} \right )^{-\frac53} \frac{1+\frac{73}{24}e^2+\frac{37}{96}e^4}{(1-e^2)^{\frac72}} \,.\label{Eq:dP_GR}
\end{equation}
with $T_\odot \equiv G M_\odot / c^3 = 4.925490947 \, \mu s $ \cite{Weisberg:2016jye} and the “tilde”-masses being equal to $\tilde m \equiv m / M_\odot$, with $M_\odot$ being the solar mass. Equivalently, defining $\delta \equiv m_g/ (\hbar \, \Omega)$, the VSL-Gravity formula is expressed as \cite{Santoni:2023uko}
\begin{equation} \label{dpdt}
    \frac{dP_{VSLG}}{dt} = -\frac{192 \pi \, T_{\odot}^{\frac53}}{5}  \frac{\tilde m_1 \tilde m_2}{\tilde M ^{\frac13}} \left (\frac{P}{2\pi} \right )^{-\frac53} \sum_{N_{min}} f(N, e, \delta, \hat n) .
\end{equation}
The sum above is performed over the Keplerian modes, $\Omega_N = N \Omega$, of the binary motion, where the function $f(N, e, \delta, \hat{n})$ is a complex combination of Bessel functions $J_N$, which involves the VSL-Gravity mass parameter $m_g$. The preferred spatial direction, $\hat{n}$, also enters this expression and can generally be parameterized by two polar angles, such that $\hat{n} = (\sin \alpha \cos \phi, \sin \alpha \sin \phi, \cos \alpha)$, with the orbital motion assumed to lie in the $x$-$y$ plane \cite{celmec}. However, since the focus of this work is on the graviton mass parameter $m_g$, we neglect the anisotropic properties of VSL-Gravity by specifically considering the simplified scenario where $\hat{n} \parallel \hat{z}$. As discussed in \cite{Santoni:2023uko}, for small values of $\delta < 1$, this configuration tends to maximize the VSL-Gravity effects, allowing us to derive the strongest possible constraints on the graviton mass. The explicit expressions for the $f$-functions in \eqref{dpdt} are included in Appendix \ref{genericfcalc} for readability.

Starting from Eq.~\eqref{dpdt}, we can calculate the Cumulative Periastron Time Shift (CPTS) and compare it with observations \cite{weisberg2004relativistic}. Given that the period time derivative \eqref{Eq:dP_dE} is generally very small in the inspiral phase of the binary ($\sim 10^{-12}$ in our case), we can neglect higher-order $P-$derivatives and terms $ O(\,[\frac{dP}{dt}]^2\,)$, leaving us with the usual leading-order approximation for the CPTS \cite{PhysRev.136.B1224,1996NCimB.111..631P}, such that

\begin{equation} \label{CPTS} 
    \Delta(t) \simeq \frac{1}{2 P}  \frac{dP}{dt} t^2  \,,
\end{equation}
where $t$ represents the time elapsed since the initial observation and $dP/dt$ is established by equations \ref{Eq:dP_GR} or \ref{dpdt}. Equation \eqref{CPTS} provides the key formula that will be used in the subsequent analysis for comparison with experimental data (Table \ref{tab:OrbDeca_data}).

\section{Methodology and Main Results}
\label{results}

\subsection{Data sets and statistical analysis}

\begin{figure}
    \centering
    \includegraphics[width=\columnwidth]{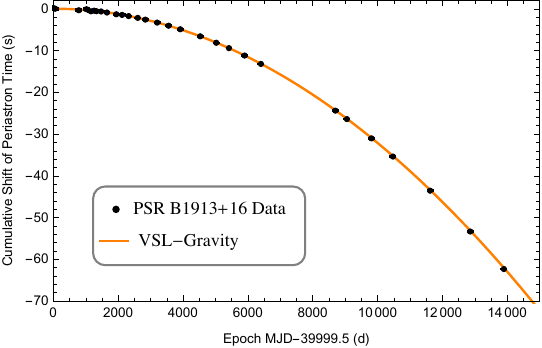}
    \caption{Cumulative shift of periastron time $\Delta(t)$  due to the loss of energy in the form of gravitational radiation in the system PSR B1913+16 for VSL-Gravity. These quantities were derived by J. M. Weisberg from data accompanying Weisberg and Huang 2016 \cite{Weisberg:2016jye}}
    \label{PSR_dat}
\end{figure}

Measurements of the cumulative periastron time shift in the PSR B1913+16 system are among the most powerful astrophysical tools to analyze how binary systems lose energy through gravitational radiation. This makes PSR B1913+16 an excellent laboratory for testing both standard and alternative gravitational theories \cite{1975ApJ...195L..51H,1982ApJ...253..908T}. This binary system consists of a neutron star and a pulsar in a highly eccentric orbit.

The dataset used in this work includes 29 measurements of the cumulative periastron time shift, $\Delta(t)$ (equation \eqref{CPTS}), spanning approximately 38.05 years, or 13,897.9 days in Modified Julian Date (MJD), which is more suitable for our analysis (see Figure \ref{PSR_dat}). These data were obtained from \cite{Weisberg:2016jye}, where J. M. Weisberg and Y. Huang conducted a relativistic analysis of 9,257 pulse time-of-arrival (TOA) measurements from PSR B1913+16, collected over the last 35 years at frequencies near 1400 MHz at the Arecibo Radio Observatory. Table \ref{tab:OrbDeca_data} presents the data used in this analysis and against which we perform our Bayesian analysis.

\begin{figure*}
    \centering
    \includegraphics[width=8.7cm]{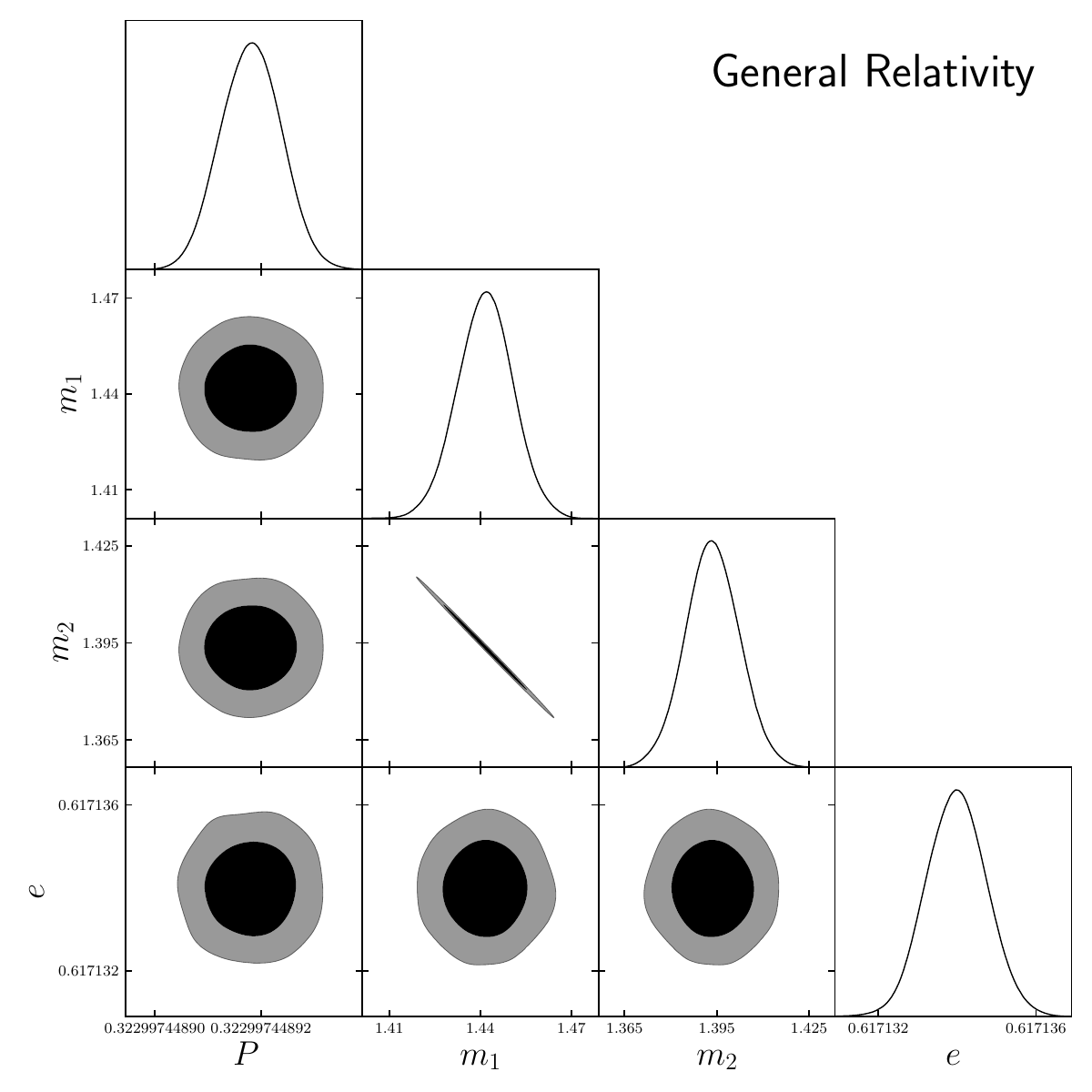} \,\,\,\,\,
    \includegraphics[width=8.7cm]{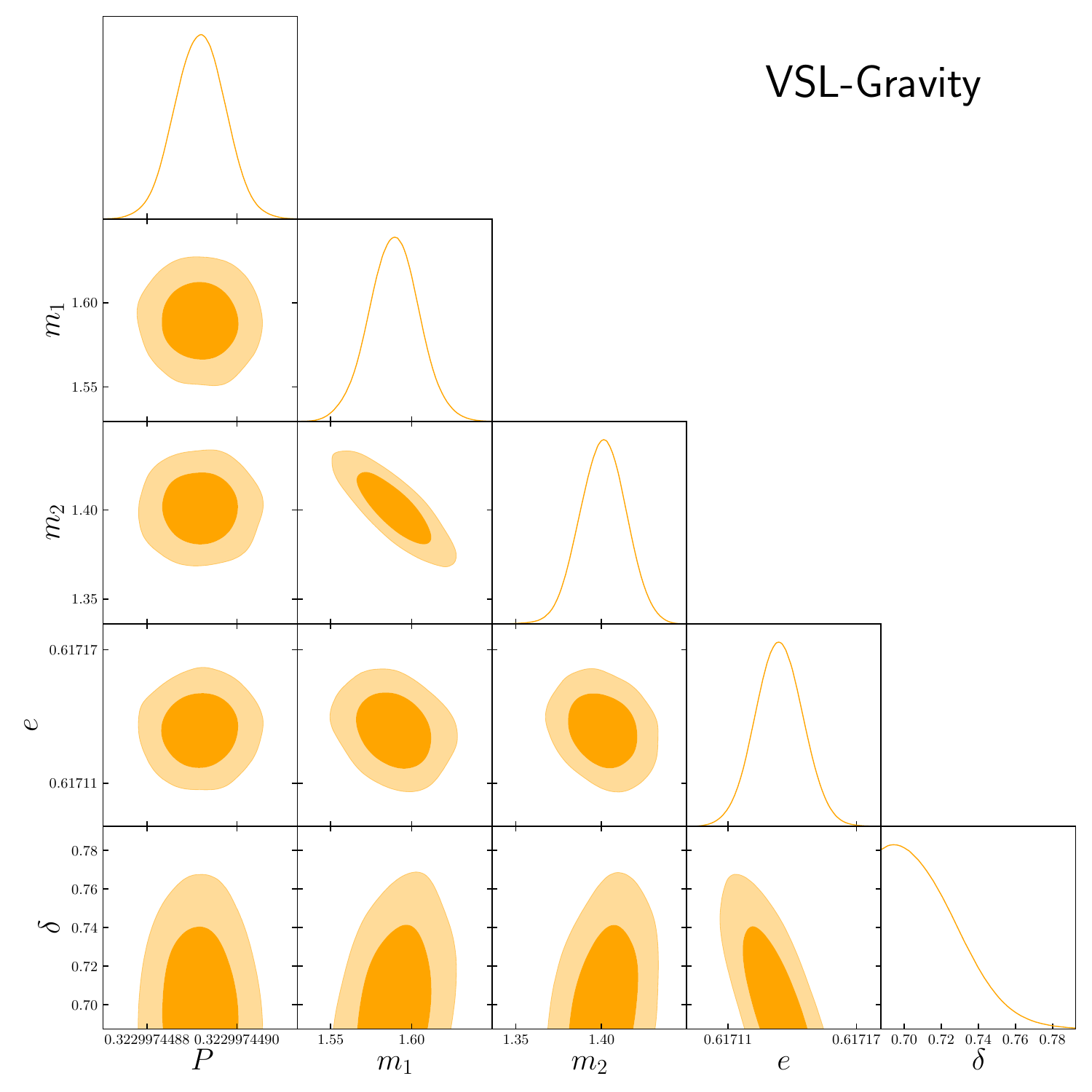}
    \caption{Two-dimensional marginalized distribution regions at 68\% and 95\% CL for the four parameters of the GR model (left panel) and for the VSL-Gravity model (right panel), including the graviton mass through the $\delta-$parameter.}
    \label{Fig:GR_VSL-Gravity_mg}
\end{figure*}

In order to constrain the free parameters in our models, we employ the Metropolis–Hastings algorithm in an appropriately modified version of \textit{emcee} sampler \cite{Foreman_Mackey_2013} to generate the chains and the GetDist Python package \cite{lewis2019getdistpythonpackageanalysing} for chain processing, ensuring that they follow the Gelman-Rubin criterion for the set of chains to be converged. In conformity with the latter, we obtain a Gelman-Rubin factor of $R-1 \lesssim 10^{-2}$. The set of scenarios considered for our statistical analysis is the following:

\begin{itemize}
    \item \textbf{GR} -- This is recovered in the massless limit $\delta \rightarrow 0$, where we have
    \[
    \lim_{\delta \rightarrow 0} f(N, e, \delta) = g(N, e) \,,
    \]
    so that the sum in \eqref{dpdt} takes the usual GR form 
    \begin{equation}\label{eq:g_Ne}
        \sum^{+\infty}_{N=1} g(N, e) = \frac{1 + \frac{73}{24} e^2 + \frac{37}{96} e^4}{(1 - e^2)^{7/2}} \,,
    \end{equation}
    as seen in Eq.\ref{Eq:dP_GR}, where $e$ is the eccentricity of the orbit in the binary system. In this case, our parameter space is given by $\theta = \left\lbrace P, m_1, m_2, e \right\rbrace$. See observational constraints in Fig. \ref{Fig:GR_VSL-Gravity_mg}.
    
    \item \textbf{VSL-Gravity} -- In particular, we specialize in the unique case occuring when $\hat{n}$ lays perpendicularly to the orbital plane, i.e. $\alpha = 0$, and the graviton is massive ($m_g \neq 0$). Then, the relevant function is $f_{//} \equiv f|_{\hat{n} \parallel \hat{z}}$ (see Eq.\ref{f//fperp}). Here, our parameter space is $\theta = \left\lbrace P, m_1, m_2, e, \delta \right\rbrace$. See observational constraints Fig. \ref{Fig:GR_VSL-Gravity_mg}.
\end{itemize}

\begin{table}[ht]
    \centering
    \begin{tabular}{c|c c c c }
    \hline
$\#$ & Epoch (Yr) & Epoch (d)  & $\Delta\, (s)$ &  $\sigma_\Delta\, (s)$\\ 

\hline \hline 

1    &  1974.78 & 2331.44613200    &   0.18    & 0.259 \\
2    &  1974.94 & 2389.58567500    &   0.00    & 0.207 \\
3    &  1976.94 & 3118.59096640    &  -0.37    & 0.168 \\
4    &  1977.59 & 3356.64010660    &  -0.12    & 0.156 \\
5    &  1977.96 & 3493.59102920    &  -0.62    & 0.097 \\
6    &  1978.24 & 3593.39725000    &  -0.47    & 0.052 \\
7    &  1978.43 & 3663.48770000    &  -0.60    & 0.156 \\
8    &  1978.82 & 3807.54457210    &  -0.66    & 0.078 \\
9    &  1979.32 & 3988.42315390    &  -0.91    & 0.065 \\
10   &  1980.11 & 4276.53689460    &  -1.33    & 0.058 \\
11   &  1980.59 & 4455.47749270    &  -1.48    & 0.058 \\
12   &  1981.15 & 4656.38191738    &  -1.76    & 0.012 \\
13   &  1981.92 & 4938.35870602    &  -2.19    & 0.010 \\
14   &  1982.56 & 5172.53186888    &  -2.61    & 0.009 \\
15   &  1983.56 & 5536.55001230    &  -3.32    & 0.009 \\
16   &  1984.52 & 5888.61724869    &  -4.07    & 0.014 \\
17   &  1985.51 & 6249.72841330    &  -4.92    & 0.034 \\
18   &  1987.19 & 6864.06958660    &  -6.61    & 0.026 \\
19   &  1988.54 & 7358.57869939    &  -8.15    & 0.012 \\
20   &  1989.61 & 7750.69761619    &  -9.44    & 0.010 \\
21   &  1990.93 & 8233.57881729    & -11.21    & 0.012 \\
22   &  1992.31 & 8736.80885642    & -13.22    & 0.021 \\
23   &  1998.60 & 11033.64375462    & -24.41    & 0.013 \\
24   &  1999.57 & 11388.61795356    & -26.44    & 0.010 \\
25   &  2001.64 & 12145.40097854    & -31.05    & 0.014 \\
26   &  2003.46 & 12808.19174194    & -35.41    & 0.004 \\
27   &  2006.60 & 13953.54068527    & -43.58    & 0.004 \\
28   &  2009.99 & 15191.26688708    & -53.36    & 0.004 \\
29   &  2012.83 &16229.38066008    & -62.35    & 0.008 \\ 
\hline \hline
    \end{tabular}
    \label{tab:OrbDeca_data}
    \caption{Measurements of Cumulative Periastron Time Shift (CPTS). Column 2. Epoch in Years, Column 3. Epoch (MJD-39999.5) (d), Column 4. Orbital Phase Shift (s), Column 5. Uncertainty ($10^{-8}$ s). These quantities were derived by J. M. Weisberg from data accompanying Weisberg and Huang 2016 \cite{Weisberg:2016jye}.}
\end{table}

We assume flat priors for the set of sampled parameters of interest, denoted as $\theta^{model}_i$, where $i$ indicates the number of free parameters estimated via Bayesian inference, such that, the common baseline parameters in the two models is taken to be: $P\in\left[ 0.1-0.4\right]$, $m_1\in\left[ 0.8-2.0\right]$, $m_2\in\left[0.8-2.0 \right]$, $e\in\left[ 0.4-0.9\right]$ and $\delta\in\left[ 0.001 -1.0\right]$. The Bayesian analysis implemented in this work utilizes either the maximum likelihood $\mathscr{L}_{\text{max}}$ or the minimum $\chi^2_{\text{min}}$, defined by the relationship:

\begin{equation}
    \mathscr{L}_{\text{max}} (\theta^m_i) = \exp \left[- \frac{1}{2} \chi^2_{\text{min}} (\theta^m_i) \right] .
\end{equation}

This approach facilitates the best fit for the parameters representing each theoretical model against the observed data points, using the chi-squared statistic:

\begin{equation}
    \chi^2_{\text{min}} (\theta^m_i) = \sum_{k = 1}^n \frac{\left( \Delta (t, \theta^m_i)_{\text{the}} - \Delta (t_k)_{\text{obs}}\right)^2 }{\sigma_k^2} \,,
\end{equation}
where $\Delta (t, \theta^m_i)_{\text{the}}$ represents the theoretical CPTS, $\Delta (t_k)_{\text{obs}}$ denotes the observed CPTS, and $\sigma_k$ is the error associated with each data point. To calculate the uncertainties, we employ the Fisher matrix formalism. The Fisher matrix coefficients, which encapsulate the Gaussian uncertainties of the parameters, are computed as:

\begin{equation}
    F_{ij} = \frac{1}{2} \frac{\partial^2 \chi^2_{\text{min}} (\theta^m_i)}{\partial \theta_i \partial \theta_j} \,,
\end{equation}
where $\left\lbrace \theta_i, \theta_j \right\rbrace$ represents the set of free parameters in each model. By inverting the Fisher matrix, we obtain the covariance matrix $\left[ F_{i,j} \right]^{-1} = \left[ C^{\text{cov}}_{i,j} \right]$, which yields the errors of the parameters and their potential correlations.

\begin{table}[ht]
    \renewcommand{\arraystretch}{1.5}
    \setlength{\tabcolsep}{8pt}
    \centering
\begin{tabular}{l|c|c}
\hline \hline
Parameter               & GR & VSL-Gravity\\
\hline
$P$ (d)                 & $0.322997448918(5)$ & $0.32299744892(1)$     \\
$ m_1 $ ($M_{\odot}$)   & $1.4415 \pm 0.0091$ & $1.589 \pm 0.029$      \\
$ m_2 $ ($M_{\odot}$)   & $1.3936 \pm 0.0087$ & $1.401 \pm 0.026$      \\
$e$                     & $0.6171340(7)$      & $0.617134(2)$          \\
$\delta$                & -                   & $<0.754$               \\
$m_g$ ($\text{eV}/c^2$) & -                   & $<1.1 \times 10^{-19}$ \\
$\chi^2_{red}$          & $2.64$              & $2.75$                 \\
\hline \hline
\end{tabular}
\caption{Constraints on the orbital parameters and graviton mass at 95\% CL. }
\label{tab:PSR_model}
\end{table}

\subsection{Results and comparison}

Table \ref{tab:PSR_model} summarizes the key results for the two models, highlighting the fit of the main free parameters and their uncertainties at the 95\% confidence level (CL). Notably, the constraints on the orbital parameters when using General Relativity (GR) align with previous studies, such as those by Weisberg and Huang \cite{Weisberg:2016jye}. We analyzed PSR B1913+16 orbital parameters in GR for two reasons: \textbf{i)} to verify that our methodology produces results consistent with other approaches. \textbf{ii)} to compare the findings between VSL-Gravity and GR, particularly regarding the inclusion of graviton mass.

In Figure \ref{Fig:GR_VSL-Gravity_mg} (top panel), we observe a strong anti-correlation between the masses of the binary system, which is not seen among the other parameters. For the VSL-Gravity model, the estimates are less precise than those for GR; however, parameters such as period and extent are consistent with GR within the uncertainty. In Figure \ref{Fig:GR_VSL-Gravity_mg} (bottom panel), we see an anti-correlation between the masses of the PSR B1913+16 binary system, with estimates of \( m_1 = 1.589 \pm 0.029 M_{\odot} \) and \( m_2 = 1.401 \pm 0.026 M_{\odot} \). These estimates highlight a significant discrepancy in the mass of the pulsar companion compared to GR results. This discrepancy arises from including the graviton mass \( m_g \) as an additional free parameter in VSL-Gravity, in contrast to the four orbital parameters considered in GR. Since we assume \( \hat{n} \) to be orthogonal to the orbital plane, the energy loss rate is independent of the additional VSR parameters. Thus, we can neglect the precession of the binary orbit (approximately \( \sim 4.2^\circ \) per revolution \cite{1995BASI...23...77S}), which would otherwise be relevant.

\begin{figure}\label{Fig:GWVelocity}
    \centering
    \includegraphics[width=8cm]{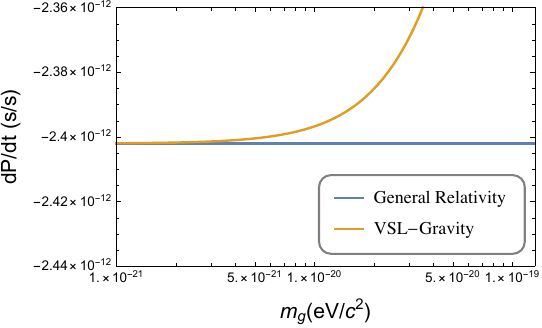}
    \caption{Comparison of the gravitational orbital decay of the binary system PSR B1913+16 in GR (equation \eqref{Eq:dP_GR}) and VSL-Gravity (equation \eqref{dpdt}).}
    \label{Fig:VSL-Gravity_mg_n}
\end{figure}

In Figure \ref{Fig:VSL-Gravity_mg_n}, we observe the rate of period change \eqref{Eq:dP_dE} as a function of the graviton mass (\( m_g \)) in the VSL-Gravity model, compared to General Relativity. For the latter, the rate remains constant to an estimated value of \( \frac{dP_{\text{GR}}}{dt} = -2.41217 \times 10^{-12} \) since Eq.\eqref{Eq:dP_GR} characterizes massless gravitons. In contrast, the curve for VSLG clearly indicates a $m_g-$dependence and is always above the GR one, implying that the energy loss due to gravitational radiation in the binary system is slightly smaller in VSL-Gravity than in GR. 

In this context, we find from our analysis for VSLG that \( \delta < 0.754 \) at 95\% CL, which bounds the graviton mass to \( m_g < 1.115 \times 10^{-19} \, \text{eV}/c^2 \) at 95\% CL.  
Several other methods have been used and studies realized in the literature to estimate constraints on the mass of the graviton, allowing for comparisons with the upper bound obtained above:

\begin{itemize}
    \item  The LIGO collaboration, using gravitational wave signals from the merger of a binary black hole in its first detection known as GW150914, achieved a strong upper bound on the graviton mass of \( m_g < 1.22 \times 10^{-22} \, \text{eV}/c^2 \) at 90\% CL \cite{LIGOScientific:2016aoc}. More recently, using information from event GW170104, the LIGO collaboration finds a bound on the mass of the graviton of \( m_g < 7.7 \times 10^{-23} \, \text{eV}/c^2 \), under the assumption that these hypothetical bosons disperse in the vacuum as massive particles \cite{LIGOScientific:2017bnn}. 
    
    \item Using data from the solar system and considering a Yukawa-like gravitational potential induced by the graviton mass (as we will see in the next section), the authors in \cite{Will:1997bb,Talmadge:1988qz} obtained an upper bound of \( m_g < 10^{-23} \, \text{eV}/c^2 \) at 90\% CL. Recently, in an updated review, Clifford derives an estimate of \( m_g < (6-10) \times 10^{-24} \, \text{eV}/c^2 \) at 90\% CL \cite{Will_2018}, while Mariani et al. determined an upper bound of \( m_g < 1.01 \times 10^{-24} \, \text{eV}/c^2 \) at 99.7\% CL using the MCMC methodology \cite{Mariani_2023}. \\
    Furthermore, as pointed out by Fienga \& Minazzoli, massive gravity theories should introduce other modifications in the dynamics of the solar system, leading to correlations between the graviton mass and the different orbital parameters, as can also be evidenced in our case using the PSR B1913+16 dataset. For a more extensive discussion see \cite{Fienga_2024,Mariani_2023}.

    \item By analyzing the trajectory of the star S2 orbiting Sagittarius A* near the galactic center, the authors in \cite{Zakharov:2018cbj} estimate \( m_g < 2.9 \times 10^{-21} \, \text{eV}/c^2 \) at 90\% CL, based on a combination of simulations and observational data. 

   \item Using the datasets of the binary pulsars PSR B1913+16 and PSR B1534+12, the authors in \cite{Finn:2001qi} find the graviton mass bound to be \( m_g < 7.6 \times 10^{-20} \, \text{eV}/c^2 \) at 90\% CL. Combining individual pulsar systems, Miao, Shao, and Ma obtain an estimate of \( m_g < 5.2 \times 10^{-21} \, \text{eV}/c^2 \) at 90\% CL, with upper bounds for the singular binaries ranging from \( m_g < (0.35 - 6.0) \times 10^{-19} \, \text{eV}/c^2 \), in agreement with the order of magnitude in the present work. For a more in-depth review of these results, see \cite{Miao:2019nhf}. Moreover, validity ranges for $m_g$ are found in \cite{Poddar:2021yjd} using binary pulsars in the context of different massive gravity realizations. For example, in the case of DGP theory, the allowed ranges are \( m_g \in (2.45 - 2.47) \times 10^{-19} \, \text{eV}/c^2 \) for PSR B1913+16 and \( m_g \in (0.31 - 1.41) \times 10^{-19} \, \text{eV}/c^2 \) for PSR J1738+0333.

    \item From the point of view of galaxy clusters, Piórkowska-Kurpas, Cao and Biesiada used data from the XMM-Newton Cluster Outskirts Project (X-COP), obtaining the constraint $m_g < (4.99 - 6.79) \times 10^{-29}\,\text{eV}$ at 95\% C.L \cite{PIORKOWSKAKURPAS202237}.  

    \item Finally, using Planck + BAO + Pantheon + KIDS-1000 data in the context of the Extended Minimal Theory of Massive Gravity (eMTMG), authors in \cite{Felice_2024} estimate an upper bound for the graviton mass of \( m_g < 6.6 \times 10^{-34} \, \text{eV}/c^2 \) at 90\% CL.

\end{itemize}

Table \ref{tab:Comparison} concisely summarizes the results of some of the cited works. Besides the comparison with alternative methodologies, it also shows the consistency of our findings with other results obtained through the same method (i.e. binary pulsar analyses). However, note that the estimates of graviton mass bounds may depend on the observation length scale, as also highlighted in \cite{Poddar:2021yjd}.

\begin{table}[ht]
    \renewcommand{\arraystretch}{1.5}
    \setlength{\tabcolsep}{8pt}
    \centering
\begin{tabular}{l|c|c|c}
\hline \hline
System/Scale            & $m_g \, (\text{eV}/c^2)$ & $\lambda_g\, (m)$ & Ref\\
\hline
Binary Pulsars    & $< 10^{-19}$             & $\gtrsim 10^{13}$ & This Work\\
                  & $<10^{-19}$              & $\gtrsim 10^{13}$ & \cite{Poddar:2021yjd}\\
                  & $<10^{-20}$              & $\gtrsim 10^{14}$ & \cite{Finn:2001qi}\\
S2 Star           & $< 10^{-21}$             & $\gtrsim 10^{15}$ & \cite{Zakharov:2018cbj}\\
GW150914          & $< 10^{-22}$             & $\gtrsim 10^{16}$ & \cite{LIGOScientific:2016aoc}\\
Solar System      & $< 10^{-24}$             & $\gtrsim 10^{17}$ &\cite{Mariani_2023} \\
Galaxy Clusters   & $< 10^{-29}$             & $\gtrsim 10^{23}$ &\cite{PIORKOWSKAKURPAS202237} \\
Cosmology         & $< 10^{-34}$             & $\gtrsim 10^{28}$ & \cite{Felice_2024} \\
\hline \hline
\end{tabular}
\caption{Comparison of the graviton mass bounds obtained in this work with respect to other methodologies and references found in the literature.}
\label{tab:Comparison}
\end{table}

\section{Theoretical consequences}
\label{consequences}

After having presented the methodology to analyze the data and the main results, in this section we will review the information obtained, focusing on the theoretical consequences of the presence of a graviton mass.
\subsection{Dispersion Relation and GW's speed}

As shown in \cite{grav3}, working in a specific gauge, VSLG leads to a Klein-Gordon equation for the $h_{\mu\nu}-$field, implying the following massive dispersion relation
\begin{equation}\label{eq:Dispersion}
(p^2 - m_g^2)h_{\mu \nu} =  0 \,.
\end{equation}
Defining the respective compton wavelength for the graviton as
\begin{equation}\label{eq:ComtomLength}
    \lambda_g = \frac{2\pi\hbar}{m_g c} \,,
\end{equation}
the “massiveness” of the gravitational oscillations results in a frequency-dependent velocity $c_{GW}(f)$ of propagating gravitational waves
\begin{equation}\label{eq:GWVelocity}
    c_{GW} = c \, \sqrt{ 1 - \left (\frac{m_g c^2}{2\pi \hbar f} \right )^2} = c\, \sqrt{1 - \left (\frac{c}{\lambda_g f} \right )^2} \,,
\end{equation}
which in turn implies a dispersion for different GW wavelengths, that has been used to constrain the graviton mass in \cite{baker2017strong,Will:1999vg}. 

In principle, the detection of the deterministic signal from event GW170817 has imposed strong limits on the speed of propagation of GWs $\vert c_{GW}/c - 1 \vert \leq 10^{-16}$, at a frequency of $f \approx 10^2$ Hz \cite{LIGOScientific:2017ync}. However, given the relation of Eq.\eqref{eq:GWVelocity}, it is expected that at low frequencies (or low energies) the GW signal propagates more slowly than at high frequencies (or high energies). In fact, as also pointed out in \cite{grav3}, the magnitude of generic VSLG effects is, in this context, related to the parameter $m_g^2/(2 \pi \hbar f)^2$, meaning that the departure from GR should be easier to detect in lower frequency ranges.

\begin{figure}[h!]
    \centering
    \includegraphics[width=8cm]{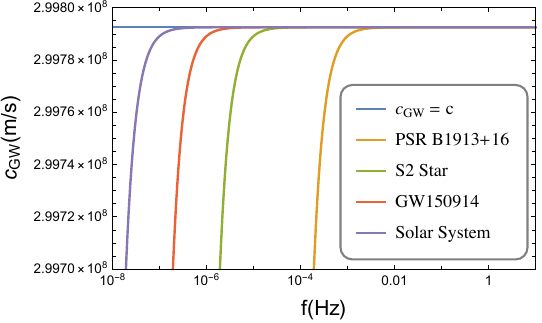}
    \caption{Frequency-dependent velocity $c_{GW}(f)$ of propagating gravitational waves, considering $m_g$ equal to the the upper bound value from various analyses. The PSR B1913+16 label refers to the present work. The horizontal axis is shown on a logarithmic scale for clarity.}
    \label{Fig:VSL-Gravity_mg_n_1}
\end{figure}

In Figure \ref{Fig:VSL-Gravity_mg_n_1} we have showcased the frequency dependence of the GW speed $c_{GW}$ using as $m_g-$value the upper limit coming from different observational sources: PSR B1913+16 (present work: $m_g \sim 1.1 \times 10^{-19} \,\text{eV}/c^2$), S2 Star at the galactic center ($m_g \sim 2.9 \times 10^{-21}\, \text{eV}/c^2 $) \cite{Zakharov:2016lzv}, GW150914 event ($m_g \sim 1.22 \times 10^{-22} \,\text{eV}/c^2$) \cite{LIGOScientific:2016aoc} and the Solar System ($m_g \sim 10^{-23}\, \text{eV}/c^2$) \cite{Will:1997bb}.

\begin{figure}[h!]
    \centering
    \includegraphics[width=8cm]{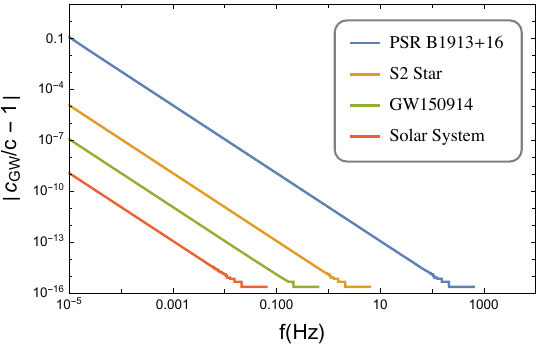}
    \caption{Proportion between the speed of propagation of gravitational and electromagnetic waves. The PSR B1913+16 tag refers to the present work. Both axis are represented in logarithmic scale for convenience.}
    \label{Fig:VSL-Gravity_mg_n_2}
\end{figure}

On the other hand, Figure \ref{Fig:VSL-Gravity_mg_n_2} shows the behavior of $\vert c_{GW}/c - 1 \vert$ for the different cases presented above (PSR B1913+16, S2 Star, GW150914 and Solar System). In the present work, at a frequency of $f \simeq 10^{-4} Hz$, we find $\vert c_{GW}/c - 1 \vert \sim 0.05$, corresponding to $c_{GW} \gtrsim 0.95 \,c$. This aligns with the forecast made in \cite{Bonilla:2019mbm} for ET and DECIGO and could also be verified in the future within the bandwidth of the LISA project \footnote{\url{https://www.lisamission.org/}}.

It is important to stress that the subliminal velocity of GWs in VSLG is directly connected to the massiveness of helicity-2 tensorial DOFs. Indeed, VSL-Gravity features a slight deformation of the two familiar plus and cross polarization states of GR. See \cite{grav3,Hou_2018,dePaula:2004bc} for a more extensive discussion.

\subsection{Yukawa potential}

In the static limit, $\partial_t h_{\mu\nu}(x) = 0$, the solution to the massive equations of motion corresponds to the standard Yukawa potential. Thus, the gravitational potential generated by a mass $M$ in this framework becomes \cite{Santoni:2023uko}
\begin{equation}
    V(r) = - \frac{G M}{r} e^{-m_g r} \,.
\end{equation}
From this expression, we can easily identify the interaction or correlation length $\lambda$ associated with this modified gravitational formulation, which coincides with the Compton wavelength of the graviton, $\lambda = \lambda_g$. Note that our upper bound for the VSLG graviton mass, $m_g \leq 10^{-19}\,\text{eV}/c^2$, can be directly translated into the following lower bound for its Compton wavelength:
\begin{equation}
    \lambda_g \gtrsim 10^{13} \,\text{m} \,.
\end{equation}
The graviton mass would have significant implications in many contexts, ranging from astrophysical systems to the large-scale structure (LSS) and dynamics of the Universe: \textbf{i}) At galactic-scale level, Zakharov et al. \cite{Zakharov:2018cbj} use data from two independent groups that track bright S-stars in the direction of Sgr A* at the center of the Milky Way to test the Yukawa gravitational potential. In particular, the precision of the astrometric data for the star S2 \cite{Gillessen_2009} is utilized, along with simulations of its orbit around the massive black hole Sgr A*, to obtain constraints on the Yukawa interaction length, specifically $\lambda \gtrsim 10^{15}\,\text{m}$ \cite{Borka_2013}. Since $\lambda = \lambda_g$, this technique provides an upper limit for the mass of the graviton around $m_g \lesssim 2.9 \times 10^{-21}\,\text{eV}/c^2$ \cite{Zakharov:2016lzv}. \textbf{ii}) It is also possible to estimate the graviton mass using the dynamical properties of galaxy clusters, associating it with the weakening of the Yukawa-like gravitational interaction as distance increases (see, for example, \cite{PIORKOWSKAKURPAS202237}). In this context, Sunyaev–Zel'dovich data and X-rays have been already used to study the profile of the gravitational field in galaxy clusters \cite{Eckert_2022}, enabling the possibility to use them as robust experimental tools to investigate the massive nature of gravity in a model-independent way. \textbf{iii}) One of the most promising explanation for the Universe's accelerated expansion, alternative to dark energy, involves a graviton Compton wavelength comparable to the Hubble length $c/H_0$, where $H_0$ is the Hubble constant. Graviton masses around $m_g \sim 10^{-33} \, \text{eV} /c^2$ would lead to a reduction in gravitational attraction at large scales, resulting in an effective accelerated expansion at the cosmological level, as observed today \cite{alves2010}. The authors in \cite{Dubovsky_2010} estimate $m_g \lesssim 4.99 \times 10^{-30} \, \text{eV}/c^2$ by studying the tensor contribution to the polarization spectrum and anisotropies in the CMB temperature in the presence of a non-zero graviton mass. One key result is the strong suppression of the CMB spectrum for large values of $m_g$. Furthermore, the authors of \cite{Jusufi_2023} propose a cosmological model based on the long-range Yukawa gravitational potential with a massive graviton, where dark matter is not explained in terms of exotic particles but rather as a consequence of the interaction between baryonic matter and this modified potential. In the latter model, dark energy emerges as a self-interaction effect among gravitons. Their estimate of the graviton mass, $m_g \simeq 10^{-33} \, \text{eV}/c^2$, leads to a value for the cosmological constant of $\Lambda \simeq 10^{-52} \, \text{m}^{-2}$. 

Other fundamental implications of massive gravity include mitigating cosmological tensions \cite{de2021minimal} and introducing novel dark matter candidates \cite{Aoki:2016zgp,PhysRevLett.128.081806}. 

\subsection{Early Universe}

Cosmic inflation has become a standard paradigm in the study of the early Universe. One of its main predictions is the generation of primordial gravitational waves (PGWs), which are usually expected to follow a nearly scale-invariant spectrum \cite{rivera2016}. PGWs are predicted to originate from quantum-mechanical perturbations of the metric during inflation, and their detection may provide additional evidence that either supports or challenges different proposals within the inflationary paradigm. One of the best ways to search for such signatures is through the measurement of the tensor-to-scalar ratio $r$, which quantifies the relative strength of tensor perturbations (related to spacetime fluctuations) compared to scalar perturbations (related to density fluctuations). Several experiments, such as Planck \cite{2020_inflation}, BICEP/Keck \cite{Tristram:2021tvh} and SKA \cite{janssen2014}, are able to constrain $r$ and allow us to obtain information on the energy scale of inflation itself. Another possibility is to use GW detectors to identify eventual PGW relics. In particular, tensorial modes with non-vanishing masses, and therefore massive gravity theories, are recognized to produce peculiar blue-tilted spectra \cite{Lin:2016gve}, enhancing the possibility of detection for higher frequencies.

\begin{figure}[h!]
    \centering
    \includegraphics[width=8.5cm]{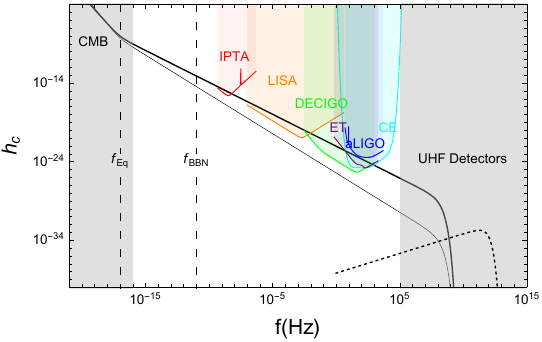}
    \caption{Characteristic strain $h_c$ produced by CGWB contributions from low to ultra high frequency, alongside the sensitivities of various detector and experiments: CMB anisotropies and polarisation (Gray band - left), \textcolor{red}{IPTA}, \textcolor{orange}{LISA}, \textcolor{green}{DECIGO}, \textcolor{violet}{ET}, \textcolor{blue}{aLIGO}, \textcolor{cyan}{CE} \& UHF Detectors (Gray band - right). The vertical dashed lines represent the characteristic frequencies of the radiation-matter equality and BBN epochs. The black solid thick curve corresponds to relics of massive PGWs (reproduced from \cite{Bartolo_2016}), while the black solid thin curve to massless ones (reproduced from \cite{Giovannini:2014jca,Giovannini_2020}). The black dotted curve is related to the thermal (massless) gravitons contribution (reproduced from \cite{Vagnozzi_2022}). Both axis are represented in logarithmic scale.}
    \label{Fig:CgB}
\end{figure}

The PGW signatures, spanning from the CMB scale or ultra-low frequencies (ULF) to ultra-high frequencies (UHF) \footnote{Ultra-High Frequency Gravitational Waves / Stephen Hawking Centre for Theoretical Cosmology: \url{https://www.ctc.cam.ac.uk/activities/UHF-GW.php}} (see, as an example, Figure \ref{Fig:CgB}), could also be detectable in the intermediate frequency range $\left[ 10^1 - 10^4 , \text{Hz}\right]$ by current interferometric detectors, such as the advanced Laser Interferometer Gravitational-Wave Observatory (aLIGO) \footnote{\url{https://advancedligo.mit.edu/}} and future ones, such as Einstein Telescope (ET) \footnote{\url{https://www.et-gw.eu/}}, Cosmic Explorer (CE) \footnote{\url{https://cosmicexplorer.org/}}, the DECi-hertz Interferometer Gravitational wave Observatory (DECIGO) \footnote{\url{https://decigo.jp/index_E.html}}. Further interesting opportunities are given by the space-based Laser Interferometer Space Antenna (LISA) \cite{Cooray:2003cv,Zhao:2009pt,Giovannini_2020} in the range $\left[ 10^{-4} - 10^0 , \text{Hz}\right]$ and, in the nHz band, by collaborations like the North American Nanohertz Observatory for Gravitational Waves (NANOGrav) \footnote{\url{https://nanograv.org/}}, the European Pulsar Timing Array (EPTA) \footnote{\url{https://www.epta.eu.org/}} and the International Pulsar Timing Array (IPTA) \footnote{\url{https://ipta4gw.org/}}. In Table \ref{tab:Comparison_1} we concisely summarize these useful experiments along with their sensitivity ranges.

\begin{table}[ht]
    \renewcommand{\arraystretch}{1.5}
    \setlength{\tabcolsep}{8pt}
    \centering
\begin{tabular}{l|c|l}
\hline \hline
Band   & Range (Hz)                        &  Detector\\
\hline
UHF    & $\left[ 10^{5}-10^{15} \right]$   &\cite{Fang_Yu_2004,2008AIPC..969.1045B}\\
HF     & $\left[ 10^1-10^{4} \right]$    & \textcolor{blue}{aLIGO}\\
       &                                   &\textcolor{green}{DECIGO}\\
       &                                   &\textcolor{violet}{ET} \\
       &                                   &\textcolor{cyan}{CE} \\
IF     & $\left[ 10^{-4}-10^{0} \right]$        & \textcolor{orange}{LISA}\\
LF     & $\left[ 10^{-9}-10^{-7} \right]$  &\textcolor{red}{NANOGrav}\\
       &                                   &\textcolor{red}{EPTA} \\
       &                                   &\textcolor{red}{IPTA} \\
ULF    & $\left[ 10^{-15}-10^{-18} \right]$ & BICEP\\
\hline \hline
\end{tabular}
\caption{Relevant detectors for CGWB measurements and their respective frequency windows, ranging several different bandwidths: \textit{Ultra High Frequency}, \textit{High Frequency}, \textit{Intermediate Frequency} and \textit{Low Frequency}.}
\label{tab:Comparison_1}
\end{table}

However, PGWs are not the only source for the cosmological gravitational wave background (CGWB) \cite{Caprini_2018,Bartolo_2016}. An additional origin is given, for example, by relic GWs generated by the decoupling of thermal gravitons from the hot primordial plasma, which is expected to occur around Planck time \cite{Giovannini:2014jca,Giovannini_2020} (see Figure \ref{Fig:CgB}). Other processes can, in principle, contribute to the CGWB by imprinting their fingerprint within the respective frequency windows. Among them, there is radiation-matter equality, characterized by frequency $f_{eq} \sim 10^{-17}\,Hz$, and neutrino decoupling, which roughly coincides with the era of the Big Bang nucleosynthesis (BBN) $f_{BBN} \sim 10^{-11}\,Hz$. Interestingly enough, we note that, as already pointed out by Vagnozzi \& Loeb in \cite{Vagnozzi_2022}, the detection of a thermally-induced component of the CGWB may have serious consequences for the inflationary paradigm itself.

At the moment, it is not fully understood how VSL-Gravity would specifically affect these cosmological scenarios because a first-principle derivation, which we plan to carry out in the future, is still missing. Nevertheless, we expect that, especially for cosmology and the early Universe, a more in-depth analysis concerning a non-linear VSLG extension might be required.

\section{Final Remarks}
\label{final}

In this paper, we established upper bounds on the graviton mass within the framework of VSL-Gravity. Using 29 data points from the Cumulative Periastron Time Shift  $\Delta(t)$ of the Hulse \& Taylor binary pulsar PSR B1913+16, we derived an estimate of \( m_g \lesssim 1.1 \times 10^{-19} \,\text{eV}/c^2 \) at 95\% CL and its associated Compton wavelength $\lambda_g \gtrsim 10^{13}$ \text{m}. In addition, we explored the consequences that graviton massiveness would have in several astrophysical and cosmological scenarios.

These findings contribute to further understanding graviton properties and their phenomenological implications, especially in the context of the VSLG theory. The search for better constraints on the graviton mass not only enhances our theoretical formulations, but also guides future experimental efforts aimed at detecting novel effects through gravitational phenomena \cite{Chung_2021}. As gravitational wave observatories advance, the insights gained from this study will prove useful in refining our search for evidence of elusive particles like gravitons, potentially unlocking new avenues in our quest for gravity quantization \cite{Tobar_2024}. 

Overall, this work underscores the importance of astrophysical observations in addressing fundamental questions about the nature of gravity and its possible quantum aspects.

\acknowledgments
The authors express their gratitude and special thanks to Joel M. Weisberg for his valuable comments and suggestions. A. Bonilla  acknowledges a fellowship (44.291/2018-0) of the PCI Program - MCTI and CNPq. A. Bonilla would also like to thank CERN-TH for the support of the “short visit" program to its facilities and professors J. R. Ellis, F. Quevedo and J. Kopp for their advice and insightful comments on this topic. A. Bonilla also wishes to thank A. Castillo for his significant contributions from a more theoretical point of view to the present project, which helped to improve the quality of the writing. A. Santoni acknowledges financial support from ANID Fellowship CONICYT-PFCHA/DoctoradoNacional/2020-21201387. RCN thanks the financial support from the Conselho Nacional de Desenvolvimento Cient\'{i}fico e Tecnologico (CNPq, National Council for Scientific and Technological Development) under the project No. 304306/2022-3, and the Fundação de Amparo à pesquisa do Estado do RS (FAPERGS, Research Support Foundation of the State of RS) for partial financial support under the project No. 23/2551-0000848-3. This article is based on work from COST Action CA21136 - “Addressing observational tensions in cosmology with systematics and fundamental physics (CosmoVerse)", supported by COST (European Cooperation in Science and Technology).

\appendix

\section{Expressions of $f(N,\delta,e,\hat n)$} \label{genericfcalc}

In this appendix, we include the explicit expressions for the function $f(N,\delta,e,\hat n)$ along with some more details. Generally, we have two contributions that can be distinguished $f_{//}$ and $f_{\perp}$
\begin{equation}\label{Eq:f_plus}
    f(N,e,\delta ,\hat n)=f_{//}(N,\delta, e )+ f_\perp(N,\delta, e ,\hat n) \, .
\end{equation}\\
Although $f_{//}$ is independent of $\hat n$ and, therefore, always present, $f_{\perp}$ depends on the preferred space direction, which means that it is relevant only when $\hat n$ is not orthogonal to the orbital plane. That also implies that in the massless graviton limit $m_g\to0$, just $f_{//}$ can and will survive. With all that being said, the expressions for both contributions are
\begin{widetext}
\begin{eqnarray} \label{f//fperp}
    f_{//}(N,e,\delta) &=& \sqrt{1-\frac{\delta^2}{N^2}} \left \{  \left [ 1 + \frac{\delta^2}{6 N^2} \left (\frac{91}{2} +\frac{16 \delta^2}{ N^2} -15 \frac{4 + \frac{3\delta^2}{N^2}}{2\sqrt{1-\frac{\delta^2}{N^2}}} \tanh^{-1}\sqrt{1-\frac{\delta^2}{N^2}} \right ) \right ] g(N,e) \right . \\
    && \;\;\;\;\;\;\;\;\;\;\;\;\;\;\;\;\;\;\;\;\;\;\;\;\;\;\;\;\;\; \left. + \frac{ \delta^2 }{48} \left (-\frac{25}{6} +\frac{80 \delta^2}{3 N^2} + 5 \frac{ 4 - \frac{15\delta^2}{N^2}}{2\sqrt{1-\frac{\delta^2}{N^2}}} \tanh^{-1}\sqrt{1-\frac{\delta^2}{N^2}}  \right ) J^2_N (Ne) \right \} \nonumber \,,\\
f_{\perp}(N, e, \delta , \hat n)&=& - \frac{5 N^2}{384} \delta ^2 \sqrt{1-\frac{\delta^2}{N^2}} \left\{  
\left (23 +\frac{16 \delta^2}{ N^2} - \frac{12 + \frac{21\delta^2}{N^2}}{\sqrt{1-\frac{\delta^2}{N^2}}} \tanh^{-1}\sqrt{1-\frac{\delta^2}{N^2}} \right ) \hat n^i \hat n^k L^{*\,ij}_N L^{kj}_N
 \right. \nonumber \\ 
&& \;\;\;\;\;\;\;\;\;\;\;\;\;\;\;\;\;\;
-\left (\frac{37}{4} +\frac{20 \delta^2}{ N^2} - \frac{12 + \frac{99\delta^2}{ N^2}}{4 \sqrt{1-\frac{\delta^2}{N^2}}} \tanh^{-1}\sqrt{1-\frac{\delta^2}{N^2}} \right )n^i \hat n^j \hat n^k \hat n^l L^{*\,ij}_N L^{kl}_N
 \nonumber\\
&& \;\;\;\;\;\;\;\;\;\;\;\;\;\;\;\;\;\;\left.
+\frac{2}{N} \left (13 -\frac{16 \delta^2}{ N^2} - \frac{12 - \frac{21\delta^2}{N^2}}{\sqrt{1-\frac{\delta^2}{N^2}}} \tanh^{-1}\sqrt{1-\frac{\delta^2}{N^2}} \right ) \sin^2\theta (\cos^2\phi \, L^{xx}_N + \sin^2 \phi \, L^{yy}_N) J_N \right \} \, , \nonumber
\end{eqnarray}
with $g(N,e)$ being the same function defined in \cite{peters1963gravitational} 
\begin{eqnarray} \label{gNe}
    g(N,e) &=& \frac{N^4}{32} \left \{ (J_{N-2} -2 e J_{N-1} + \frac2N J_N +2 e J_{N+1}-J_{N+2})^2 + (1-e^2)(J_{N-2} -2 J_N +J_{N+2} \,)^2 +\frac{4}{3N^2} J^2_N \right \} \,, \nonumber \\
    \end{eqnarray}
being $N e$ the argument of the above $J-$functions, and
\begin{eqnarray}
    \hat n^i \hat n^j L^{*\,ik}_N L^{jk}_N &=& \sin^2\theta (\cos^2\phi \, (L^{xx}_N)^2 + \sin^2 \phi \,(L^{yy}_N)^2) 
     \;\;\;+ \sin^2\theta |L^{xy}_N|^2 \,, \nonumber\\
    n^i \hat n^j \hat n^k \hat n^l L^{*\,ij}_N L^{kl}_N &=& \sin^4\theta (\cos^2\phi \, L^{xx}_N + \sin^2 \phi\, L^{yy}_N)^2\
     \;\;\;+4 \sin^4 \theta \sin^2\phi \cos^2\phi |L^{xy}_N|^2 \,. \nonumber \\
\end{eqnarray}
\end{widetext}
%such that, in the limit of $\delta = 0$, we can recover the equation \ref{eq:g_Ne} as it should be in the case of GR.

%\textit{LISA} (top-left), \textit{aLIGO} (top-right), \textit{ET} (bottom-left) and \textit{DECIGO} (bottom-right)

%The \nocite command causes all entries in a bibliography to be printed out whether or not they are actually referenced in the text. This is appropriate for the sample file to show the different styles of references, but authors most likely will not want to use it.

\bibliography{apssamp,Bbinary,Bmassgrav}% Produces the bibliography via BibTeX.

\providecommand{\noopsort}[1]{}\providecommand{\singleletter}[1]{#1}%\providecommand{\noopsort}[1]{}\providecommand{\singleletter}[1]{#1}%\providecommand{\noopsort}[1]{}\providecommand{\singleletter}[1]{#1}%
\begin{thebibliography}{86}%
\makeatletter
\providecommand \@ifxundefined [1]{%
 \@ifx{#1\undefined}
}%
\providecommand \@ifnum [1]{%
 \ifnum #1\expandafter \@firstoftwo
 \else \expandafter \@secondoftwo
 \fi
}%
\providecommand \@ifx [1]{%
 \ifx #1\expandafter \@firstoftwo
 \else \expandafter \@secondoftwo
 \fi
}%
\providecommand \natexlab [1]{#1}%
\providecommand \enquote  [1]{``#1''}%
\providecommand \bibnamefont  [1]{#1}%
\providecommand \bibfnamefont [1]{#1}%
\providecommand \citenamefont [1]{#1}%
\providecommand \href@noop [0]{\@secondoftwo}%
\providecommand \href [0]{\begingroup \@sanitize@url \@href}%
\providecommand \@href[1]{\@@startlink{#1}\@@href}%
\providecommand \@@href[1]{\endgroup#1\@@endlink}%
\providecommand \@sanitize@url [0]{\catcode `\\12\catcode `\$12\catcode `\&12\catcode `\#12\catcode `\^12\catcode `\_12\catcode `\%12\relax}%
\providecommand \@@startlink[1]{}%
\providecommand \@@endlink[0]{}%
\providecommand \url  [0]{\begingroup\@sanitize@url \@url }%
\providecommand \@url [1]{\endgroup\@href {#1}{\urlprefix }}%
\providecommand \urlprefix  [0]{URL }%
\providecommand \Eprint [0]{\href }%
\providecommand \doibase [0]{https://doi.org/}%
\providecommand \selectlanguage [0]{\@gobble}%
\providecommand \bibinfo  [0]{\@secondoftwo}%
\providecommand \bibfield  [0]{\@secondoftwo}%
\providecommand \translation [1]{[#1]}%
\providecommand \BibitemOpen [0]{}%
\providecommand \bibitemStop [0]{}%
\providecommand \bibitemNoStop [0]{.\EOS\space}%
\providecommand \EOS [0]{\spacefactor3000\relax}%
\providecommand \BibitemShut  [1]{\csname bibitem#1\endcsname}%
\let\auto@bib@innerbib\@empty
%</preamble>
\bibitem [{\citenamefont {Fierz}\ and\ \citenamefont {Pauli}(1939)}]{Fierz:1939ix}%
  \BibitemOpen
  \bibfield  {author} {\bibinfo {author} {\bibfnamefont {M.}~\bibnamefont {Fierz}}\ and\ \bibinfo {author} {\bibfnamefont {W.}~\bibnamefont {Pauli}},\ }\bibfield  {title} {\bibinfo {title} {{On relativistic wave equations for particles of arbitrary spin in an electromagnetic field}},\ }\href {https://doi.org/10.1098/rspa.1939.0140} {\bibfield  {journal} {\bibinfo  {journal} {Proc. Roy. Soc. Lond. A}\ }\textbf {\bibinfo {volume} {173}},\ \bibinfo {pages} {211} (\bibinfo {year} {1939})}\BibitemShut {NoStop}%
\bibitem [{\citenamefont {van Dam}\ and\ \citenamefont {Veltman}(1970)}]{vanDam:1970vg}%
  \BibitemOpen
  \bibfield  {author} {\bibinfo {author} {\bibfnamefont {H.}~\bibnamefont {van Dam}}\ and\ \bibinfo {author} {\bibfnamefont {M.~J.~G.}\ \bibnamefont {Veltman}},\ }\bibfield  {title} {\bibinfo {title} {{Massive and massless Yang-Mills and gravitational fields}},\ }\href {https://doi.org/10.1016/0550-3213(70)90416-5} {\bibfield  {journal} {\bibinfo  {journal} {Nucl. Phys. B}\ }\textbf {\bibinfo {volume} {22}},\ \bibinfo {pages} {397} (\bibinfo {year} {1970})}\BibitemShut {NoStop}%
\bibitem [{\citenamefont {Zakharov}(1970)}]{Zakharov:1970cc}%
  \BibitemOpen
  \bibfield  {author} {\bibinfo {author} {\bibfnamefont {V.~I.}\ \bibnamefont {Zakharov}},\ }\bibfield  {title} {\bibinfo {title} {{Linearized gravitation theory and the graviton mass}},\ }\href@noop {} {\bibfield  {journal} {\bibinfo  {journal} {JETP Lett.}\ }\textbf {\bibinfo {volume} {12}},\ \bibinfo {pages} {312} (\bibinfo {year} {1970})}\BibitemShut {NoStop}%
\bibitem [{\citenamefont {Hinterbichler}(2012)}]{hinterbichler2012theoretical}%
  \BibitemOpen
  \bibfield  {author} {\bibinfo {author} {\bibfnamefont {K.}~\bibnamefont {Hinterbichler}},\ }\bibfield  {title} {\bibinfo {title} {{Theoretical Aspects of Massive Gravity}},\ }\href {https://doi.org/10.1103/RevModPhys.84.671} {\bibfield  {journal} {\bibinfo  {journal} {Rev. Mod. Phys.}\ }\textbf {\bibinfo {volume} {84}},\ \bibinfo {pages} {671} (\bibinfo {year} {2012})},\ \Eprint {https://arxiv.org/abs/1105.3735} {arXiv:1105.3735 [hep-th]} \BibitemShut {NoStop}%
\bibitem [{\citenamefont {de~Araujo}\ \emph {et~al.}(2021)\citenamefont {de~Araujo}, \citenamefont {De~Felice}, \citenamefont {Kumar},\ and\ \citenamefont {Nunes}}]{de2021minimal}%
  \BibitemOpen
  \bibfield  {author} {\bibinfo {author} {\bibfnamefont {J.~C.~N.}\ \bibnamefont {de~Araujo}}, \bibinfo {author} {\bibfnamefont {A.}~\bibnamefont {De~Felice}}, \bibinfo {author} {\bibfnamefont {S.}~\bibnamefont {Kumar}},\ and\ \bibinfo {author} {\bibfnamefont {R.~C.}\ \bibnamefont {Nunes}},\ }\bibfield  {title} {\bibinfo {title} {{Minimal theory of massive gravity in the light of CMB data and the S8 tension}},\ }\href {https://doi.org/10.1103/PhysRevD.104.104057} {\bibfield  {journal} {\bibinfo  {journal} {Phys. Rev. D}\ }\textbf {\bibinfo {volume} {104}},\ \bibinfo {pages} {104057} (\bibinfo {year} {2021})},\ \Eprint {https://arxiv.org/abs/2106.09595} {arXiv:2106.09595 [astro-ph.CO]} \BibitemShut {NoStop}%
\bibitem [{\citenamefont {Vainshtein}(1972)}]{Vainshtein:1972sx}%
  \BibitemOpen
  \bibfield  {author} {\bibinfo {author} {\bibfnamefont {A.~I.}\ \bibnamefont {Vainshtein}},\ }\bibfield  {title} {\bibinfo {title} {{To the problem of nonvanishing gravitation mass}},\ }\href {https://doi.org/10.1016/0370-2693(72)90147-5} {\bibfield  {journal} {\bibinfo  {journal} {Phys. Lett. B}\ }\textbf {\bibinfo {volume} {39}},\ \bibinfo {pages} {393} (\bibinfo {year} {1972})}\BibitemShut {NoStop}%
\bibitem [{\citenamefont {Dvali}\ \emph {et~al.}(2000{\natexlab{a}})\citenamefont {Dvali}, \citenamefont {Gabadadze},\ and\ \citenamefont {Porrati}}]{Dvali:2000hr}%
  \BibitemOpen
  \bibfield  {author} {\bibinfo {author} {\bibfnamefont {G.~R.}\ \bibnamefont {Dvali}}, \bibinfo {author} {\bibfnamefont {G.}~\bibnamefont {Gabadadze}},\ and\ \bibinfo {author} {\bibfnamefont {M.}~\bibnamefont {Porrati}},\ }\bibfield  {title} {\bibinfo {title} {{4-D gravity on a brane in 5-D Minkowski space}},\ }\href {https://doi.org/10.1016/S0370-2693(00)00669-9} {\bibfield  {journal} {\bibinfo  {journal} {Phys. Lett. B}\ }\textbf {\bibinfo {volume} {485}},\ \bibinfo {pages} {208} (\bibinfo {year} {2000}{\natexlab{a}})},\ \Eprint {https://arxiv.org/abs/hep-th/0005016} {arXiv:hep-th/0005016} \BibitemShut {NoStop}%
\bibitem [{\citenamefont {Dvali}\ \emph {et~al.}(2000{\natexlab{b}})\citenamefont {Dvali}, \citenamefont {Gabadadze},\ and\ \citenamefont {Porrati}}]{Dvali:2000rv}%
  \BibitemOpen
  \bibfield  {author} {\bibinfo {author} {\bibfnamefont {G.~R.}\ \bibnamefont {Dvali}}, \bibinfo {author} {\bibfnamefont {G.}~\bibnamefont {Gabadadze}},\ and\ \bibinfo {author} {\bibfnamefont {M.}~\bibnamefont {Porrati}},\ }\bibfield  {title} {\bibinfo {title} {{Metastable gravitons and infinite volume extra dimensions}},\ }\href {https://doi.org/10.1016/S0370-2693(00)00631-6} {\bibfield  {journal} {\bibinfo  {journal} {Phys. Lett. B}\ }\textbf {\bibinfo {volume} {484}},\ \bibinfo {pages} {112} (\bibinfo {year} {2000}{\natexlab{b}})},\ \Eprint {https://arxiv.org/abs/hep-th/0002190} {arXiv:hep-th/0002190} \BibitemShut {NoStop}%
\bibitem [{\citenamefont {de~Rham}\ \emph {et~al.}(2011)\citenamefont {de~Rham}, \citenamefont {Gabadadze},\ and\ \citenamefont {Tolley}}]{deRham:2010kj}%
  \BibitemOpen
  \bibfield  {author} {\bibinfo {author} {\bibfnamefont {C.}~\bibnamefont {de~Rham}}, \bibinfo {author} {\bibfnamefont {G.}~\bibnamefont {Gabadadze}},\ and\ \bibinfo {author} {\bibfnamefont {A.~J.}\ \bibnamefont {Tolley}},\ }\bibfield  {title} {\bibinfo {title} {{Resummation of Massive Gravity}},\ }\href {https://doi.org/10.1103/PhysRevLett.106.231101} {\bibfield  {journal} {\bibinfo  {journal} {Phys. Rev. Lett.}\ }\textbf {\bibinfo {volume} {106}},\ \bibinfo {pages} {231101} (\bibinfo {year} {2011})},\ \Eprint {https://arxiv.org/abs/1011.1232} {arXiv:1011.1232 [hep-th]} \BibitemShut {NoStop}%
\bibitem [{\citenamefont {Alfaro}\ and\ \citenamefont {Santoni}(2022)}]{grav3}%
  \BibitemOpen
  \bibfield  {author} {\bibinfo {author} {\bibfnamefont {J.}~\bibnamefont {Alfaro}}\ and\ \bibinfo {author} {\bibfnamefont {A.}~\bibnamefont {Santoni}},\ }\bibfield  {title} {\bibinfo {title} {{Very special linear gravity: A gauge-invariant graviton mass}},\ }\href {https://doi.org/10.1016/j.physletb.2022.137080} {\bibfield  {journal} {\bibinfo  {journal} {Phys. Lett. B}\ }\textbf {\bibinfo {volume} {829}},\ \bibinfo {pages} {137080} (\bibinfo {year} {2022})},\ \Eprint {https://arxiv.org/abs/2204.05485} {arXiv:2204.05485 [gr-qc]} \BibitemShut {NoStop}%
\bibitem [{\citenamefont {Cohen}\ and\ \citenamefont {Glashow}(2006{\natexlab{a}})}]{vsr1}%
  \BibitemOpen
  \bibfield  {author} {\bibinfo {author} {\bibfnamefont {A.~G.}\ \bibnamefont {Cohen}}\ and\ \bibinfo {author} {\bibfnamefont {S.~L.}\ \bibnamefont {Glashow}},\ }\bibfield  {title} {\bibinfo {title} {{Very special relativity}},\ }\href {https://doi.org/10.1103/PhysRevLett.97.021601} {\bibfield  {journal} {\bibinfo  {journal} {Phys. Rev. Lett.}\ }\textbf {\bibinfo {volume} {97}},\ \bibinfo {pages} {021601} (\bibinfo {year} {2006}{\natexlab{a}})},\ \Eprint {https://arxiv.org/abs/hep-ph/0601236} {arXiv:hep-ph/0601236} \BibitemShut {NoStop}%
\bibitem [{\citenamefont {Cohen}\ and\ \citenamefont {Glashow}(2006{\natexlab{b}})}]{vsr2}%
  \BibitemOpen
  \bibfield  {author} {\bibinfo {author} {\bibfnamefont {A.~G.}\ \bibnamefont {Cohen}}\ and\ \bibinfo {author} {\bibfnamefont {S.~L.}\ \bibnamefont {Glashow}},\ }\bibfield  {title} {\bibinfo {title} {{A Lorentz-Violating Origin of Neutrino Mass?}},\ }\href@noop {} {\bibfield  {journal} {\bibinfo  {journal} {arXiv preprint}\ } (\bibinfo {year} {2006}{\natexlab{b}})},\ \Eprint {https://arxiv.org/abs/hep-ph/0605036} {arXiv:hep-ph/0605036} \BibitemShut {NoStop}%
\bibitem [{\citenamefont {Alfaro}\ and\ \citenamefont {Soto}(2019)}]{vsrqed}%
  \BibitemOpen
  \bibfield  {author} {\bibinfo {author} {\bibfnamefont {J.}~\bibnamefont {Alfaro}}\ and\ \bibinfo {author} {\bibfnamefont {A.}~\bibnamefont {Soto}},\ }\bibfield  {title} {\bibinfo {title} {Photon mass in very special relativity},\ }\href@noop {} {\bibfield  {journal} {\bibinfo  {journal} {Physical Review D}\ }\textbf {\bibinfo {volume} {100}},\ \bibinfo {pages} {055029} (\bibinfo {year} {2019})}\BibitemShut {NoStop}%
\bibitem [{\citenamefont {Alfaro}\ and\ \citenamefont {Rivelles}(2013)}]{Alfaro:2013uva}%
  \BibitemOpen
  \bibfield  {author} {\bibinfo {author} {\bibfnamefont {J.}~\bibnamefont {Alfaro}}\ and\ \bibinfo {author} {\bibfnamefont {V.~O.}\ \bibnamefont {Rivelles}},\ }\bibfield  {title} {\bibinfo {title} {{Non Abelian Fields in Very Special Relativity}},\ }\href {https://doi.org/10.1103/PhysRevD.88.085023} {\bibfield  {journal} {\bibinfo  {journal} {Phys. Rev. D}\ }\textbf {\bibinfo {volume} {88}},\ \bibinfo {pages} {085023} (\bibinfo {year} {2013})},\ \Eprint {https://arxiv.org/abs/1305.1577} {arXiv:1305.1577 [hep-th]} \BibitemShut {NoStop}%
\bibitem [{\citenamefont {Santoni}(2024)}]{Santoni:2024coa}%
  \BibitemOpen
  \bibfield  {author} {\bibinfo {author} {\bibfnamefont {A.}~\bibnamefont {Santoni}},\ }\emph {\bibinfo {title} {{Delving into the phenomenology of very special relativity: From subatomic particles to binary stars}}},\ \href {https://doi.org/10.34726/hss.2024.124925} {Ph.D. thesis} (\bibinfo {year} {2024}),\ \Eprint {https://arxiv.org/abs/2409.03104} {arXiv:2409.03104 [hep-ph]} \BibitemShut {NoStop}%
\bibitem [{\citenamefont {Giovannini}(2002)}]{Giovannini:2002sv}%
  \BibitemOpen
  \bibfield  {author} {\bibinfo {author} {\bibfnamefont {M.}~\bibnamefont {Giovannini}},\ }\bibfield  {title} {\bibinfo {title} {{Primordial magnetic fields}},\ }in\ \href@noop {} {\emph {\bibinfo {booktitle} {{7th Paris Cosmology Colloquium on High Energy Astrophysics for and from Space}}}}\ (\bibinfo {year} {2002})\ \Eprint {https://arxiv.org/abs/hep-ph/0208152} {arXiv:hep-ph/0208152} \BibitemShut {NoStop}%
\bibitem [{\citenamefont {Ilderton}(2016)}]{background}%
  \BibitemOpen
  \bibfield  {author} {\bibinfo {author} {\bibfnamefont {A.}~\bibnamefont {Ilderton}},\ }\bibfield  {title} {\bibinfo {title} {Very special relativity as a background field theory},\ }\href@noop {} {\bibfield  {journal} {\bibinfo  {journal} {Physical Review D}\ }\textbf {\bibinfo {volume} {94}},\ \bibinfo {pages} {045019} (\bibinfo {year} {2016})}\BibitemShut {NoStop}%
\bibitem [{\citenamefont {Das}\ and\ \citenamefont {Faizal}(2018)}]{Das:2018umm}%
  \BibitemOpen
  \bibfield  {author} {\bibinfo {author} {\bibfnamefont {S.}~\bibnamefont {Das}}\ and\ \bibinfo {author} {\bibfnamefont {M.}~\bibnamefont {Faizal}},\ }\bibfield  {title} {\bibinfo {title} {{Dimensional reduction via a novel Higgs mechanism}},\ }\href {https://doi.org/10.1007/s10714-018-2409-x} {\bibfield  {journal} {\bibinfo  {journal} {Gen. Rel. Grav.}\ }\textbf {\bibinfo {volume} {50}},\ \bibinfo {pages} {87} (\bibinfo {year} {2018})},\ \Eprint {https://arxiv.org/abs/1806.07520} {arXiv:1806.07520 [gr-qc]} \BibitemShut {NoStop}%
\bibitem [{\citenamefont {Mann}\ \emph {et~al.}(2021)\citenamefont {Mann}, \citenamefont {Husin}, \citenamefont {Patel}, \citenamefont {Faizal}, \citenamefont {Sulaksono},\ and\ \citenamefont {Suroso}}]{Mann:2020jcu}%
  \BibitemOpen
  \bibfield  {author} {\bibinfo {author} {\bibfnamefont {R.~B.}\ \bibnamefont {Mann}}, \bibinfo {author} {\bibfnamefont {I.}~\bibnamefont {Husin}}, \bibinfo {author} {\bibfnamefont {H.}~\bibnamefont {Patel}}, \bibinfo {author} {\bibfnamefont {M.}~\bibnamefont {Faizal}}, \bibinfo {author} {\bibfnamefont {A.}~\bibnamefont {Sulaksono}},\ and\ \bibinfo {author} {\bibfnamefont {A.}~\bibnamefont {Suroso}},\ }\bibfield  {title} {\bibinfo {title} {{Testing Short Distance Anisotropy in Space}},\ }\href {https://doi.org/10.1038/s41598-021-86355-3} {\bibfield  {journal} {\bibinfo  {journal} {Sci. Rep.}\ }\textbf {\bibinfo {volume} {11}},\ \bibinfo {pages} {7474} (\bibinfo {year} {2021})},\ \Eprint {https://arxiv.org/abs/2011.03340} {arXiv:2011.03340 [gr-qc]} \BibitemShut {NoStop}%
\bibitem [{\citenamefont {{Hulse}}\ and\ \citenamefont {{Taylor}}(1975)}]{1975ApJ...195L..51H}%
  \BibitemOpen
  \bibfield  {author} {\bibinfo {author} {\bibfnamefont {R.~A.}\ \bibnamefont {{Hulse}}}\ and\ \bibinfo {author} {\bibfnamefont {J.~H.}\ \bibnamefont {{Taylor}}},\ }\bibfield  {title} {\bibinfo {title} {{Discovery of a pulsar in a binary system.}},\ }\href {https://doi.org/10.1086/181708} {\bibfield  {journal} {\bibinfo  {journal} {Astrophys. J Lett.}\ }\textbf {\bibinfo {volume} {195}},\ \bibinfo {pages} {L51} (\bibinfo {year} {1975})}\BibitemShut {NoStop}%
\bibitem [{\citenamefont {{Taylor}}\ and\ \citenamefont {{Weisberg}}(1982)}]{1982ApJ...253..908T}%
  \BibitemOpen
  \bibfield  {author} {\bibinfo {author} {\bibfnamefont {J.~H.}\ \bibnamefont {{Taylor}}}\ and\ \bibinfo {author} {\bibfnamefont {J.~M.}\ \bibnamefont {{Weisberg}}},\ }\bibfield  {title} {\bibinfo {title} {{A new test of general relativity - Gravitational radiation and the binary pulsar PSR 1913+16}},\ }\href {https://doi.org/10.1086/159690} {\bibfield  {journal} {\bibinfo  {journal} {\apj}\ }\textbf {\bibinfo {volume} {253}},\ \bibinfo {pages} {908} (\bibinfo {year} {1982})}\BibitemShut {NoStop}%
\bibitem [{\citenamefont {Poddar}\ \emph {et~al.}(2022)\citenamefont {Poddar}, \citenamefont {Mohanty},\ and\ \citenamefont {Jana}}]{Poddar:2021yjd}%
  \BibitemOpen
  \bibfield  {author} {\bibinfo {author} {\bibfnamefont {T.~K.}\ \bibnamefont {Poddar}}, \bibinfo {author} {\bibfnamefont {S.}~\bibnamefont {Mohanty}},\ and\ \bibinfo {author} {\bibfnamefont {S.}~\bibnamefont {Jana}},\ }\bibfield  {title} {\bibinfo {title} {{Gravitational radiation from binary systems in massive graviton theories}},\ }\href {https://doi.org/10.1088/1475-7516/2022/03/019} {\bibfield  {journal} {\bibinfo  {journal} {JCAP}\ }\textbf {\bibinfo {volume} {03}},\ \bibinfo {pages} {019}},\ \Eprint {https://arxiv.org/abs/2105.13335} {arXiv:2105.13335 [gr-qc]} \BibitemShut {NoStop}%
\bibitem [{\citenamefont {Shao}\ \emph {et~al.}(2020)\citenamefont {Shao}, \citenamefont {Wex},\ and\ \citenamefont {Zhou}}]{shao2020new}%
  \BibitemOpen
  \bibfield  {author} {\bibinfo {author} {\bibfnamefont {L.}~\bibnamefont {Shao}}, \bibinfo {author} {\bibfnamefont {N.}~\bibnamefont {Wex}},\ and\ \bibinfo {author} {\bibfnamefont {S.-Y.}\ \bibnamefont {Zhou}},\ }\bibfield  {title} {\bibinfo {title} {{New graviton mass bound from binary pulsars}},\ }\href@noop {} {\bibfield  {journal} {\bibinfo  {journal} {Phys. Rev. D}\ }\textbf {\bibinfo {volume} {102}},\ \bibinfo {pages} {024069} (\bibinfo {year} {2020})}\BibitemShut {NoStop}%
\bibitem [{\citenamefont {Weisberg}\ \emph {et~al.}(2010)\citenamefont {Weisberg}, \citenamefont {Nice},\ and\ \citenamefont {Taylor}}]{Weisberg:2010zz}%
  \BibitemOpen
  \bibfield  {author} {\bibinfo {author} {\bibfnamefont {J.~M.}\ \bibnamefont {Weisberg}}, \bibinfo {author} {\bibfnamefont {D.~J.}\ \bibnamefont {Nice}},\ and\ \bibinfo {author} {\bibfnamefont {J.~H.}\ \bibnamefont {Taylor}},\ }\bibfield  {title} {\bibinfo {title} {{Timing Measurements of the Relativistic Binary Pulsar PSR B1913+16}},\ }\href {https://doi.org/10.1088/0004-637X/722/2/1030} {\bibfield  {journal} {\bibinfo  {journal} {Astrophys. J.}\ }\textbf {\bibinfo {volume} {722}},\ \bibinfo {pages} {1030} (\bibinfo {year} {2010})},\ \Eprint {https://arxiv.org/abs/1011.0718} {arXiv:1011.0718 [astro-ph.GA]} \BibitemShut {NoStop}%
\bibitem [{\citenamefont {Santoni}\ \emph {et~al.}(2023)\citenamefont {Santoni}, \citenamefont {Alfaro},\ and\ \citenamefont {Soto}}]{Santoni:2023uko}%
  \BibitemOpen
  \bibfield  {author} {\bibinfo {author} {\bibfnamefont {A.}~\bibnamefont {Santoni}}, \bibinfo {author} {\bibfnamefont {J.}~\bibnamefont {Alfaro}},\ and\ \bibinfo {author} {\bibfnamefont {A.}~\bibnamefont {Soto}},\ }\bibfield  {title} {\bibinfo {title} {{Graviton mass bounds in very special relativity from binary pulsar\textquoteright{}s gravitational waves}},\ }\href {https://doi.org/10.1103/PhysRevD.108.044072} {\bibfield  {journal} {\bibinfo  {journal} {Phys. Rev. D}\ }\textbf {\bibinfo {volume} {108}},\ \bibinfo {pages} {044072} (\bibinfo {year} {2023})},\ \Eprint {https://arxiv.org/abs/2306.02464} {arXiv:2306.02464 [gr-qc]} \BibitemShut {NoStop}%
\bibitem [{\citenamefont {Peters}\ and\ \citenamefont {Mathews}(1963)}]{peters1963gravitational}%
  \BibitemOpen
  \bibfield  {author} {\bibinfo {author} {\bibfnamefont {P.~C.}\ \bibnamefont {Peters}}\ and\ \bibinfo {author} {\bibfnamefont {J.}~\bibnamefont {Mathews}},\ }\bibfield  {title} {\bibinfo {title} {{Gravitational radiation from point masses in a Keplerian orbit}},\ }\href@noop {} {\bibfield  {journal} {\bibinfo  {journal} {Phys. Rev.}\ }\textbf {\bibinfo {volume} {131}},\ \bibinfo {pages} {435} (\bibinfo {year} {1963})}\BibitemShut {NoStop}%
\bibitem [{\citenamefont {Weisberg}\ and\ \citenamefont {Huang}(2016)}]{Weisberg:2016jye}%
  \BibitemOpen
  \bibfield  {author} {\bibinfo {author} {\bibfnamefont {J.~M.}\ \bibnamefont {Weisberg}}\ and\ \bibinfo {author} {\bibfnamefont {Y.}~\bibnamefont {Huang}},\ }\bibfield  {title} {\bibinfo {title} {{Relativistic Measurements from Timing the Binary Pulsar PSR B1913+16}},\ }\href {https://doi.org/10.3847/0004-637X/829/1/55} {\bibfield  {journal} {\bibinfo  {journal} {Astrophys. J.}\ }\textbf {\bibinfo {volume} {829}},\ \bibinfo {pages} {55} (\bibinfo {year} {2016})},\ \Eprint {https://arxiv.org/abs/1606.02744} {arXiv:1606.02744 [astro-ph.HE]} \BibitemShut {NoStop}%
\bibitem [{\citenamefont {Klioner}(2016)}]{celmec}%
  \BibitemOpen
  \bibfield  {author} {\bibinfo {author} {\bibfnamefont {S.~A.}\ \bibnamefont {Klioner}},\ }\bibfield  {title} {\bibinfo {title} {Basic celestial mechanics},\ }\href@noop {} {\bibfield  {journal} {\bibinfo  {journal} {arXiv preprint arXiv:1609.00915}\ } (\bibinfo {year} {2016})}\BibitemShut {NoStop}%
\bibitem [{\citenamefont {Weisberg}\ and\ \citenamefont {Taylor}(2004)}]{weisberg2004relativistic}%
  \BibitemOpen
  \bibfield  {author} {\bibinfo {author} {\bibfnamefont {J.~M.}\ \bibnamefont {Weisberg}}\ and\ \bibinfo {author} {\bibfnamefont {J.~H.}\ \bibnamefont {Taylor}},\ }\bibfield  {title} {\bibinfo {title} {Relativistic binary pulsar b1913+ 16: Thirty years of observations and analysis},\ }\href@noop {} {\bibfield  {journal} {\bibinfo  {journal} {arXiv preprint astro-ph/0407149}\ } (\bibinfo {year} {2004})}\BibitemShut {NoStop}%
\bibitem [{\citenamefont {Peters}(1964)}]{PhysRev.136.B1224}%
  \BibitemOpen
  \bibfield  {author} {\bibinfo {author} {\bibfnamefont {P.~C.}\ \bibnamefont {Peters}},\ }\bibfield  {title} {\bibinfo {title} {Gravitational radiation and the motion of two point masses},\ }\href {https://doi.org/10.1103/PhysRev.136.B1224} {\bibfield  {journal} {\bibinfo  {journal} {Phys. Rev.}\ }\textbf {\bibinfo {volume} {136}},\ \bibinfo {pages} {B1224} (\bibinfo {year} {1964})}\BibitemShut {NoStop}%
\bibitem [{\citenamefont {{Pierro}}\ and\ \citenamefont {{Pinto}}(1996)}]{1996NCimB.111..631P}%
  \BibitemOpen
  \bibfield  {author} {\bibinfo {author} {\bibfnamefont {V.}~\bibnamefont {{Pierro}}}\ and\ \bibinfo {author} {\bibfnamefont {I.~M.}\ \bibnamefont {{Pinto}}},\ }\bibfield  {title} {\bibinfo {title} {{Exact solution of Peters-Mathews equations for any orbital eccentricity.}},\ }\href {https://doi.org/10.1007/BF02726655} {\bibfield  {journal} {\bibinfo  {journal} {Nuovo Cimento B Serie}\ }\textbf {\bibinfo {volume} {111B}},\ \bibinfo {pages} {631} (\bibinfo {year} {1996})}\BibitemShut {NoStop}%
\bibitem [{\citenamefont {Foreman-Mackey}\ \emph {et~al.}(2013)\citenamefont {Foreman-Mackey}, \citenamefont {Hogg}, \citenamefont {Lang},\ and\ \citenamefont {Goodman}}]{Foreman_Mackey_2013}%
  \BibitemOpen
  \bibfield  {author} {\bibinfo {author} {\bibfnamefont {D.}~\bibnamefont {Foreman-Mackey}}, \bibinfo {author} {\bibfnamefont {D.~W.}\ \bibnamefont {Hogg}}, \bibinfo {author} {\bibfnamefont {D.}~\bibnamefont {Lang}},\ and\ \bibinfo {author} {\bibfnamefont {J.}~\bibnamefont {Goodman}},\ }\bibfield  {title} {\bibinfo {title} {<tt>emcee</tt>: The mcmc hammer},\ }\href {https://doi.org/10.1086/670067} {\bibfield  {journal} {\bibinfo  {journal} {Publications of the Astronomical Society of the Pacific}\ }\textbf {\bibinfo {volume} {125}},\ \bibinfo {pages} {306–312} (\bibinfo {year} {2013})}\BibitemShut {NoStop}%
\bibitem [{\citenamefont {Lewis}(2019)}]{lewis2019getdistpythonpackageanalysing}%
  \BibitemOpen
  \bibfield  {author} {\bibinfo {author} {\bibfnamefont {A.}~\bibnamefont {Lewis}},\ }\href {https://arxiv.org/abs/1910.13970} {\bibinfo {title} {Getdist: a python package for analysing monte carlo samples}} (\bibinfo {year} {2019}),\ \Eprint {https://arxiv.org/abs/1910.13970} {arXiv:1910.13970 [astro-ph.IM]} \BibitemShut {NoStop}%
\bibitem [{\citenamefont {{Sivaram}}(1995)}]{1995BASI...23...77S}%
  \BibitemOpen
  \bibfield  {author} {\bibinfo {author} {\bibfnamefont {C.}~\bibnamefont {{Sivaram}}},\ }\bibfield  {title} {\bibinfo {title} {{The Hulse-Taylor binary pulsar PSR 1913+16.}},\ }\href@noop {} {\bibfield  {journal} {\bibinfo  {journal} {Bulletin of the Astronomical Society of India}\ }\textbf {\bibinfo {volume} {23}},\ \bibinfo {pages} {77} (\bibinfo {year} {1995})}\BibitemShut {NoStop}%
\bibitem [{\citenamefont {Abbott}\ \emph {et~al.}(2016)\citenamefont {Abbott} \emph {et~al.}}]{LIGOScientific:2016aoc}%
  \BibitemOpen
  \bibfield  {author} {\bibinfo {author} {\bibfnamefont {B.~P.}\ \bibnamefont {Abbott}} \emph {et~al.} (\bibinfo {collaboration} {LIGO Scientific, Virgo}),\ }\bibfield  {title} {\bibinfo {title} {{Observation of Gravitational Waves from a Binary Black Hole Merger}},\ }\href {https://doi.org/10.1103/PhysRevLett.116.061102} {\bibfield  {journal} {\bibinfo  {journal} {Phys. Rev. Lett.}\ }\textbf {\bibinfo {volume} {116}},\ \bibinfo {pages} {061102} (\bibinfo {year} {2016})},\ \Eprint {https://arxiv.org/abs/1602.03837} {arXiv:1602.03837 [gr-qc]} \BibitemShut {NoStop}%
\bibitem [{\citenamefont {Abbott}\ \emph {et~al.}(2017{\natexlab{a}})\citenamefont {Abbott} \emph {et~al.}}]{LIGOScientific:2017bnn}%
  \BibitemOpen
  \bibfield  {author} {\bibinfo {author} {\bibfnamefont {B.~P.}\ \bibnamefont {Abbott}} \emph {et~al.} (\bibinfo {collaboration} {LIGO Scientific, VIRGO}),\ }\bibfield  {title} {\bibinfo {title} {{GW170104: Observation of a 50-Solar-Mass Binary Black Hole Coalescence at Redshift 0.2}},\ }\href {https://doi.org/10.1103/PhysRevLett.118.221101} {\bibfield  {journal} {\bibinfo  {journal} {Phys. Rev. Lett.}\ }\textbf {\bibinfo {volume} {118}},\ \bibinfo {pages} {221101} (\bibinfo {year} {2017}{\natexlab{a}})},\ \bibinfo {note} {[Erratum: Phys.Rev.Lett. 121, 129901 (2018)]},\ \Eprint {https://arxiv.org/abs/1706.01812} {arXiv:1706.01812 [gr-qc]} \BibitemShut {NoStop}%
\bibitem [{\citenamefont {Will}(1998)}]{Will:1997bb}%
  \BibitemOpen
  \bibfield  {author} {\bibinfo {author} {\bibfnamefont {C.~M.}\ \bibnamefont {Will}},\ }\bibfield  {title} {\bibinfo {title} {{Bounding the mass of the graviton using gravitational wave observations of inspiralling compact binaries}},\ }\href {https://doi.org/10.1103/PhysRevD.57.2061} {\bibfield  {journal} {\bibinfo  {journal} {Phys. Rev. D}\ }\textbf {\bibinfo {volume} {57}},\ \bibinfo {pages} {2061} (\bibinfo {year} {1998})},\ \Eprint {https://arxiv.org/abs/gr-qc/9709011} {arXiv:gr-qc/9709011} \BibitemShut {NoStop}%
\bibitem [{\citenamefont {Talmadge}\ \emph {et~al.}(1988)\citenamefont {Talmadge}, \citenamefont {Berthias}, \citenamefont {Hellings},\ and\ \citenamefont {Standish}}]{Talmadge:1988qz}%
  \BibitemOpen
  \bibfield  {author} {\bibinfo {author} {\bibfnamefont {C.}~\bibnamefont {Talmadge}}, \bibinfo {author} {\bibfnamefont {J.~P.}\ \bibnamefont {Berthias}}, \bibinfo {author} {\bibfnamefont {R.~W.}\ \bibnamefont {Hellings}},\ and\ \bibinfo {author} {\bibfnamefont {E.~M.}\ \bibnamefont {Standish}},\ }\bibfield  {title} {\bibinfo {title} {{Model Independent Constraints on Possible Modifications of Newtonian Gravity}},\ }\href {https://doi.org/10.1103/PhysRevLett.61.1159} {\bibfield  {journal} {\bibinfo  {journal} {Phys. Rev. Lett.}\ }\textbf {\bibinfo {volume} {61}},\ \bibinfo {pages} {1159} (\bibinfo {year} {1988})}\BibitemShut {NoStop}%
\bibitem [{\citenamefont {Will}(2018)}]{Will_2018}%
  \BibitemOpen
  \bibfield  {author} {\bibinfo {author} {\bibfnamefont {C.~M.}\ \bibnamefont {Will}},\ }\bibfield  {title} {\bibinfo {title} {Solar system versus gravitational-wave bounds on the graviton mass},\ }\href {https://doi.org/10.1088/1361-6382/aad13c} {\bibfield  {journal} {\bibinfo  {journal} {Classical and Quantum Gravity}\ }\textbf {\bibinfo {volume} {35}},\ \bibinfo {pages} {17LT01} (\bibinfo {year} {2018})}\BibitemShut {NoStop}%
\bibitem [{\citenamefont {Mariani}\ \emph {et~al.}(2023)\citenamefont {Mariani}, \citenamefont {Fienga}, \citenamefont {Minazzoli}, \citenamefont {Gastineau},\ and\ \citenamefont {Laskar}}]{Mariani_2023}%
  \BibitemOpen
  \bibfield  {author} {\bibinfo {author} {\bibfnamefont {V.}~\bibnamefont {Mariani}}, \bibinfo {author} {\bibfnamefont {A.}~\bibnamefont {Fienga}}, \bibinfo {author} {\bibfnamefont {O.}~\bibnamefont {Minazzoli}}, \bibinfo {author} {\bibfnamefont {M.}~\bibnamefont {Gastineau}},\ and\ \bibinfo {author} {\bibfnamefont {J.}~\bibnamefont {Laskar}},\ }\bibfield  {title} {\bibinfo {title} {Bayesian test of the mass of the graviton with planetary ephemerides},\ }\bibfield  {journal} {\bibinfo  {journal} {Physical Review D}\ }\textbf {\bibinfo {volume} {108}},\ \href {https://doi.org/10.1103/physrevd.108.024047} {10.1103/physrevd.108.024047} (\bibinfo {year} {2023})\BibitemShut {NoStop}%
\bibitem [{\citenamefont {Fienga}\ and\ \citenamefont {Minazzoli}(2024)}]{Fienga_2024}%
  \BibitemOpen
  \bibfield  {author} {\bibinfo {author} {\bibfnamefont {A.}~\bibnamefont {Fienga}}\ and\ \bibinfo {author} {\bibfnamefont {O.}~\bibnamefont {Minazzoli}},\ }\bibfield  {title} {\bibinfo {title} {Testing theories of gravity with planetary ephemerides},\ }\bibfield  {journal} {\bibinfo  {journal} {Living Reviews in Relativity}\ }\textbf {\bibinfo {volume} {27}},\ \href {https://doi.org/10.1007/s41114-023-00047-0} {10.1007/s41114-023-00047-0} (\bibinfo {year} {2024})\BibitemShut {NoStop}%
\bibitem [{\citenamefont {Zakharov}\ \emph {et~al.}(2018)\citenamefont {Zakharov}, \citenamefont {Jovanovi\'c}, \citenamefont {Borka},\ and\ \citenamefont {Borka~Jovanovi\'c}}]{Zakharov:2018cbj}%
  \BibitemOpen
  \bibfield  {author} {\bibinfo {author} {\bibfnamefont {A.~F.}\ \bibnamefont {Zakharov}}, \bibinfo {author} {\bibfnamefont {P.}~\bibnamefont {Jovanovi\'c}}, \bibinfo {author} {\bibfnamefont {D.}~\bibnamefont {Borka}},\ and\ \bibinfo {author} {\bibfnamefont {V.}~\bibnamefont {Borka~Jovanovi\'c}},\ }\bibfield  {title} {\bibinfo {title} {{Constraining the range of Yukawa gravity interaction from S2 star orbits III: improvement expectations for graviton mass bounds}},\ }\href {https://doi.org/10.1088/1475-7516/2018/04/050} {\bibfield  {journal} {\bibinfo  {journal} {JCAP}\ }\textbf {\bibinfo {volume} {04}},\ \bibinfo {pages} {050}},\ \Eprint {https://arxiv.org/abs/1801.04679} {arXiv:1801.04679 [gr-qc]} \BibitemShut {NoStop}%
\bibitem [{\citenamefont {Finn}\ and\ \citenamefont {Sutton}(2002)}]{Finn:2001qi}%
  \BibitemOpen
  \bibfield  {author} {\bibinfo {author} {\bibfnamefont {L.~S.}\ \bibnamefont {Finn}}\ and\ \bibinfo {author} {\bibfnamefont {P.~J.}\ \bibnamefont {Sutton}},\ }\bibfield  {title} {\bibinfo {title} {{Bounding the mass of the graviton using binary pulsar observations}},\ }\href {https://doi.org/10.1103/PhysRevD.65.044022} {\bibfield  {journal} {\bibinfo  {journal} {Phys. Rev. D}\ }\textbf {\bibinfo {volume} {65}},\ \bibinfo {pages} {044022} (\bibinfo {year} {2002})},\ \Eprint {https://arxiv.org/abs/gr-qc/0109049} {arXiv:gr-qc/0109049} \BibitemShut {NoStop}%
\bibitem [{\citenamefont {Miao}\ \emph {et~al.}(2019)\citenamefont {Miao}, \citenamefont {Shao},\ and\ \citenamefont {Ma}}]{Miao:2019nhf}%
  \BibitemOpen
  \bibfield  {author} {\bibinfo {author} {\bibfnamefont {X.}~\bibnamefont {Miao}}, \bibinfo {author} {\bibfnamefont {L.}~\bibnamefont {Shao}},\ and\ \bibinfo {author} {\bibfnamefont {B.-Q.}\ \bibnamefont {Ma}},\ }\bibfield  {title} {\bibinfo {title} {{Bounding the mass of graviton in a dynamic regime with binary pulsars}},\ }\href {https://doi.org/10.1103/PhysRevD.99.123015} {\bibfield  {journal} {\bibinfo  {journal} {Phys. Rev. D}\ }\textbf {\bibinfo {volume} {99}},\ \bibinfo {pages} {123015} (\bibinfo {year} {2019})},\ \Eprint {https://arxiv.org/abs/1905.12836} {arXiv:1905.12836 [astro-ph.CO]} \BibitemShut {NoStop}%
\bibitem [{\citenamefont {Piórkowska-Kurpas}\ \emph {et~al.}(2022)\citenamefont {Piórkowska-Kurpas}, \citenamefont {Cao},\ and\ \citenamefont {Biesiada}}]{PIORKOWSKAKURPAS202237}%
  \BibitemOpen
  \bibfield  {author} {\bibinfo {author} {\bibfnamefont {A.}~\bibnamefont {Piórkowska-Kurpas}}, \bibinfo {author} {\bibfnamefont {S.}~\bibnamefont {Cao}},\ and\ \bibinfo {author} {\bibfnamefont {M.}~\bibnamefont {Biesiada}},\ }\bibfield  {title} {\bibinfo {title} {Graviton mass from x-cop galaxy clusters},\ }\href {https://doi.org/https://doi.org/10.1016/j.jheap.2022.01.001} {\bibfield  {journal} {\bibinfo  {journal} {Journal of High Energy Astrophysics}\ }\textbf {\bibinfo {volume} {33}},\ \bibinfo {pages} {37} (\bibinfo {year} {2022})}\BibitemShut {NoStop}%
\bibitem [{\citenamefont {Felice}\ \emph {et~al.}(2024)\citenamefont {Felice}, \citenamefont {Kumar}, \citenamefont {Mukohyama},\ and\ \citenamefont {Nunes}}]{Felice_2024}%
  \BibitemOpen
  \bibfield  {author} {\bibinfo {author} {\bibfnamefont {A.~D.}\ \bibnamefont {Felice}}, \bibinfo {author} {\bibfnamefont {S.}~\bibnamefont {Kumar}}, \bibinfo {author} {\bibfnamefont {S.}~\bibnamefont {Mukohyama}},\ and\ \bibinfo {author} {\bibfnamefont {R.~C.}\ \bibnamefont {Nunes}},\ }\bibfield  {title} {\bibinfo {title} {Observational bounds on extended minimal theories of massive gravity: new limits on the graviton mass},\ }\href {https://doi.org/10.1088/1475-7516/2024/04/013} {\bibfield  {journal} {\bibinfo  {journal} {Journal of Cosmology and Astroparticle Physics}\ }\textbf {\bibinfo {volume} {2024}}\bibinfo  {number} { (04)},\ \bibinfo {pages} {013}}\BibitemShut {NoStop}%
\bibitem [{\citenamefont {Baker}\ \emph {et~al.}(2017)\citenamefont {Baker}, \citenamefont {Bellini}, \citenamefont {Ferreira}, \citenamefont {Lagos}, \citenamefont {Noller},\ and\ \citenamefont {Sawicki}}]{baker2017strong}%
  \BibitemOpen
\bibfield  {number} {  }\bibfield  {author} {\bibinfo {author} {\bibfnamefont {T.}~\bibnamefont {Baker}}, \bibinfo {author} {\bibfnamefont {E.}~\bibnamefont {Bellini}}, \bibinfo {author} {\bibfnamefont {P.~G.}\ \bibnamefont {Ferreira}}, \bibinfo {author} {\bibfnamefont {M.}~\bibnamefont {Lagos}}, \bibinfo {author} {\bibfnamefont {J.}~\bibnamefont {Noller}},\ and\ \bibinfo {author} {\bibfnamefont {I.}~\bibnamefont {Sawicki}},\ }\bibfield  {title} {\bibinfo {title} {{Strong constraints on cosmological gravity from GW170817 and GRB 170817A}},\ }\href@noop {} {\bibfield  {journal} {\bibinfo  {journal} {Phys. Rev. Lett.}\ }\textbf {\bibinfo {volume} {119}},\ \bibinfo {pages} {251301} (\bibinfo {year} {2017})}\BibitemShut {NoStop}%
\bibitem [{\citenamefont {Will}(1999)}]{Will:1999vg}%
  \BibitemOpen
  \bibfield  {author} {\bibinfo {author} {\bibfnamefont {C.~M.}\ \bibnamefont {Will}},\ }\bibfield  {title} {\bibinfo {title} {{Gravitational radiation and the validity of general relativity}},\ }\href {https://doi.org/10.1063/1.882860} {\bibfield  {journal} {\bibinfo  {journal} {Phys. Today}\ }\textbf {\bibinfo {volume} {52N10}},\ \bibinfo {pages} {38} (\bibinfo {year} {1999})}\BibitemShut {NoStop}%
\bibitem [{\citenamefont {Abbott}\ \emph {et~al.}(2017{\natexlab{b}})\citenamefont {Abbott} \emph {et~al.}}]{LIGOScientific:2017ync}%
  \BibitemOpen
  \bibfield  {author} {\bibinfo {author} {\bibfnamefont {B.~P.}\ \bibnamefont {Abbott}} \emph {et~al.},\ }\bibfield  {title} {\bibinfo {title} {{Multi-messenger Observations of a Binary Neutron Star Merger}},\ }\href {https://doi.org/10.3847/2041-8213/aa91c9} {\bibfield  {journal} {\bibinfo  {journal} {Astrophys. J. Lett.}\ }\textbf {\bibinfo {volume} {848}},\ \bibinfo {pages} {L12} (\bibinfo {year} {2017}{\natexlab{b}})},\ \Eprint {https://arxiv.org/abs/1710.05833} {arXiv:1710.05833 [astro-ph.HE]} \BibitemShut {NoStop}%
\bibitem [{\citenamefont {Zakharov}\ \emph {et~al.}(2016)\citenamefont {Zakharov}, \citenamefont {Jovanovic}, \citenamefont {Borka},\ and\ \citenamefont {Jovanovic}}]{Zakharov:2016lzv}%
  \BibitemOpen
  \bibfield  {author} {\bibinfo {author} {\bibfnamefont {A.~F.}\ \bibnamefont {Zakharov}}, \bibinfo {author} {\bibfnamefont {P.}~\bibnamefont {Jovanovic}}, \bibinfo {author} {\bibfnamefont {D.}~\bibnamefont {Borka}},\ and\ \bibinfo {author} {\bibfnamefont {V.~B.}\ \bibnamefont {Jovanovic}},\ }\bibfield  {title} {\bibinfo {title} {{Constraining the range of Yukawa gravity interaction from S2 star orbits II: Bounds on graviton mass}},\ }\href {https://doi.org/10.1088/1475-7516/2016/05/045} {\bibfield  {journal} {\bibinfo  {journal} {JCAP}\ }\textbf {\bibinfo {volume} {05}},\ \bibinfo {pages} {045}},\ \Eprint {https://arxiv.org/abs/1605.00913} {arXiv:1605.00913 [gr-qc]} \BibitemShut {NoStop}%
\bibitem [{\citenamefont {Bonilla}\ \emph {et~al.}(2020)\citenamefont {Bonilla}, \citenamefont {D'Agostino}, \citenamefont {Nunes},\ and\ \citenamefont {de~Araujo}}]{Bonilla:2019mbm}%
  \BibitemOpen
  \bibfield  {author} {\bibinfo {author} {\bibfnamefont {A.}~\bibnamefont {Bonilla}}, \bibinfo {author} {\bibfnamefont {R.}~\bibnamefont {D'Agostino}}, \bibinfo {author} {\bibfnamefont {R.~C.}\ \bibnamefont {Nunes}},\ and\ \bibinfo {author} {\bibfnamefont {J.~C.~N.}\ \bibnamefont {de~Araujo}},\ }\bibfield  {title} {\bibinfo {title} {{Forecasts on the speed of gravitational waves at high $z$}},\ }\href {https://doi.org/10.1088/1475-7516/2020/03/015} {\bibfield  {journal} {\bibinfo  {journal} {JCAP}\ }\textbf {\bibinfo {volume} {03}},\ \bibinfo {pages} {015}},\ \Eprint {https://arxiv.org/abs/1910.05631} {arXiv:1910.05631 [gr-qc]} \BibitemShut {NoStop}%
\bibitem [{Note1()}]{Note1}%
  \BibitemOpen
  \bibinfo {note} {\protect \url {https://www.lisamission.org/}}\BibitemShut {NoStop}%
\bibitem [{\citenamefont {Hou}\ \emph {et~al.}(2018)\citenamefont {Hou}, \citenamefont {Gong},\ and\ \citenamefont {Liu}}]{Hou_2018}%
  \BibitemOpen
  \bibfield  {author} {\bibinfo {author} {\bibfnamefont {S.}~\bibnamefont {Hou}}, \bibinfo {author} {\bibfnamefont {Y.}~\bibnamefont {Gong}},\ and\ \bibinfo {author} {\bibfnamefont {Y.}~\bibnamefont {Liu}},\ }\bibfield  {title} {\bibinfo {title} {Polarizations of gravitational waves in horndeski theory},\ }\bibfield  {journal} {\bibinfo  {journal} {The European Physical Journal C}\ }\textbf {\bibinfo {volume} {78}},\ \href {https://doi.org/10.1140/epjc/s10052-018-5869-y} {10.1140/epjc/s10052-018-5869-y} (\bibinfo {year} {2018})\BibitemShut {NoStop}%
\bibitem [{\citenamefont {de~Paula}\ \emph {et~al.}(2004)\citenamefont {de~Paula}, \citenamefont {Miranda},\ and\ \citenamefont {Marinho}}]{dePaula:2004bc}%
  \BibitemOpen
  \bibfield  {author} {\bibinfo {author} {\bibfnamefont {W.~L.~S.}\ \bibnamefont {de~Paula}}, \bibinfo {author} {\bibfnamefont {O.~D.}\ \bibnamefont {Miranda}},\ and\ \bibinfo {author} {\bibfnamefont {R.~M.}\ \bibnamefont {Marinho}},\ }\bibfield  {title} {\bibinfo {title} {{Polarization states of gravitational waves with a massive graviton}},\ }\href {https://doi.org/10.1088/0264-9381/21/19/008} {\bibfield  {journal} {\bibinfo  {journal} {Class. Quant. Grav.}\ }\textbf {\bibinfo {volume} {21}},\ \bibinfo {pages} {4595} (\bibinfo {year} {2004})},\ \Eprint {https://arxiv.org/abs/gr-qc/0409041} {arXiv:gr-qc/0409041} \BibitemShut {NoStop}%
\bibitem [{\citenamefont {Gillessen}\ \emph {et~al.}(2009)\citenamefont {Gillessen}, \citenamefont {Eisenhauer}, \citenamefont {Fritz}, \citenamefont {Bartko}, \citenamefont {Dodds-Eden}, \citenamefont {Pfuhl}, \citenamefont {Ott},\ and\ \citenamefont {Genzel}}]{Gillessen_2009}%
  \BibitemOpen
  \bibfield  {author} {\bibinfo {author} {\bibfnamefont {S.}~\bibnamefont {Gillessen}}, \bibinfo {author} {\bibfnamefont {F.}~\bibnamefont {Eisenhauer}}, \bibinfo {author} {\bibfnamefont {T.~K.}\ \bibnamefont {Fritz}}, \bibinfo {author} {\bibfnamefont {H.}~\bibnamefont {Bartko}}, \bibinfo {author} {\bibfnamefont {K.}~\bibnamefont {Dodds-Eden}}, \bibinfo {author} {\bibfnamefont {O.}~\bibnamefont {Pfuhl}}, \bibinfo {author} {\bibfnamefont {T.}~\bibnamefont {Ott}},\ and\ \bibinfo {author} {\bibfnamefont {R.}~\bibnamefont {Genzel}},\ }\bibfield  {title} {\bibinfo {title} {The orbit of the star s2 around sgr a* from very large telescope and keck data},\ }\href {https://doi.org/10.1088/0004-637x/707/2/l114} {\bibfield  {journal} {\bibinfo  {journal} {The Astrophysical Journal}\ }\textbf {\bibinfo {volume} {707}},\ \bibinfo {pages} {L114–L117} (\bibinfo {year} {2009})}\BibitemShut {NoStop}%
\bibitem [{\citenamefont {Borka}\ \emph {et~al.}(2013)\citenamefont {Borka}, \citenamefont {Jovanović}, \citenamefont {Jovanović},\ and\ \citenamefont {Zakharov}}]{Borka_2013}%
  \BibitemOpen
  \bibfield  {author} {\bibinfo {author} {\bibfnamefont {D.}~\bibnamefont {Borka}}, \bibinfo {author} {\bibfnamefont {P.}~\bibnamefont {Jovanović}}, \bibinfo {author} {\bibfnamefont {V.~B.}\ \bibnamefont {Jovanović}},\ and\ \bibinfo {author} {\bibfnamefont {A.}~\bibnamefont {Zakharov}},\ }\bibfield  {title} {\bibinfo {title} {Constraining the range of yukawa gravity interaction from s2 star orbits},\ }\href {https://doi.org/10.1088/1475-7516/2013/11/050} {\bibfield  {journal} {\bibinfo  {journal} {Journal of Cosmology and Astroparticle Physics}\ }\textbf {\bibinfo {volume} {2013}}\bibinfo  {number} { (11)},\ \bibinfo {pages} {050–050}}\BibitemShut {NoStop}%
\bibitem [{\citenamefont {Eckert}\ \emph {et~al.}(2022)\citenamefont {Eckert}, \citenamefont {Ettori}, \citenamefont {Pointecouteau}, \citenamefont {van~der Burg},\ and\ \citenamefont {Loubser}}]{Eckert_2022}%
  \BibitemOpen
\bibfield  {number} {  }\bibfield  {author} {\bibinfo {author} {\bibfnamefont {D.}~\bibnamefont {Eckert}}, \bibinfo {author} {\bibfnamefont {S.}~\bibnamefont {Ettori}}, \bibinfo {author} {\bibfnamefont {E.}~\bibnamefont {Pointecouteau}}, \bibinfo {author} {\bibfnamefont {R.~F.~J.}\ \bibnamefont {van~der Burg}},\ and\ \bibinfo {author} {\bibfnamefont {S.~I.}\ \bibnamefont {Loubser}},\ }\bibfield  {title} {\bibinfo {title} {The gravitational field of x-cop galaxy clusters},\ }\href {https://doi.org/10.1051/0004-6361/202142507} {\bibfield  {journal} {\bibinfo  {journal} {Astron. Astrophys.}\ }\textbf {\bibinfo {volume} {662}},\ \bibinfo {pages} {A123} (\bibinfo {year} {2022})}\BibitemShut {NoStop}%
\bibitem [{\citenamefont {Alves}\ \emph {et~al.}(2010)\citenamefont {Alves}, \citenamefont {Miranda},\ and\ \citenamefont {de~Araujo}}]{alves2010}%
  \BibitemOpen
  \bibfield  {author} {\bibinfo {author} {\bibfnamefont {M.~E.~S.}\ \bibnamefont {Alves}}, \bibinfo {author} {\bibfnamefont {O.~D.}\ \bibnamefont {Miranda}},\ and\ \bibinfo {author} {\bibfnamefont {J.~C.~N.}\ \bibnamefont {de~Araujo}},\ }\href {https://arxiv.org/abs/0907.5190} {\bibinfo {title} {Can massive gravitons be an alternative to dark energy?}} (\bibinfo {year} {2010}),\ \Eprint {https://arxiv.org/abs/0907.5190} {arXiv:0907.5190 [astro-ph.CO]} \BibitemShut {NoStop}%
\bibitem [{\citenamefont {Dubovsky}\ \emph {et~al.}(2010)\citenamefont {Dubovsky}, \citenamefont {Flauger}, \citenamefont {Starobinsky},\ and\ \citenamefont {Tkachev}}]{Dubovsky_2010}%
  \BibitemOpen
  \bibfield  {author} {\bibinfo {author} {\bibfnamefont {S.}~\bibnamefont {Dubovsky}}, \bibinfo {author} {\bibfnamefont {R.}~\bibnamefont {Flauger}}, \bibinfo {author} {\bibfnamefont {A.}~\bibnamefont {Starobinsky}},\ and\ \bibinfo {author} {\bibfnamefont {I.}~\bibnamefont {Tkachev}},\ }\bibfield  {title} {\bibinfo {title} {Signatures of a graviton mass in the cosmic microwave background},\ }\bibfield  {journal} {\bibinfo  {journal} {Physical Review D}\ }\textbf {\bibinfo {volume} {81}},\ \href {https://doi.org/10.1103/physrevd.81.023523} {10.1103/physrevd.81.023523} (\bibinfo {year} {2010})\BibitemShut {NoStop}%
\bibitem [{\citenamefont {Jusufi}\ \emph {et~al.}(2023)\citenamefont {Jusufi}, \citenamefont {Leon},\ and\ \citenamefont {Millano}}]{Jusufi_2023}%
  \BibitemOpen
  \bibfield  {author} {\bibinfo {author} {\bibfnamefont {K.}~\bibnamefont {Jusufi}}, \bibinfo {author} {\bibfnamefont {G.}~\bibnamefont {Leon}},\ and\ \bibinfo {author} {\bibfnamefont {A.~D.}\ \bibnamefont {Millano}},\ }\bibfield  {title} {\bibinfo {title} {Dark universe phenomenology from yukawa potential?},\ }\href {https://doi.org/10.1016/j.dark.2023.101318} {\bibfield  {journal} {\bibinfo  {journal} {Physics of the Dark Universe}\ }\textbf {\bibinfo {volume} {42}},\ \bibinfo {pages} {101318} (\bibinfo {year} {2023})}\BibitemShut {NoStop}%
\bibitem [{\citenamefont {Aoki}\ and\ \citenamefont {Mukohyama}(2016)}]{Aoki:2016zgp}%
  \BibitemOpen
  \bibfield  {author} {\bibinfo {author} {\bibfnamefont {K.}~\bibnamefont {Aoki}}\ and\ \bibinfo {author} {\bibfnamefont {S.}~\bibnamefont {Mukohyama}},\ }\bibfield  {title} {\bibinfo {title} {{Massive gravitons as dark matter and gravitational waves}},\ }\href {https://doi.org/10.1103/PhysRevD.94.024001} {\bibfield  {journal} {\bibinfo  {journal} {Phys. Rev. D}\ }\textbf {\bibinfo {volume} {94}},\ \bibinfo {pages} {024001} (\bibinfo {year} {2016})},\ \Eprint {https://arxiv.org/abs/1604.06704} {arXiv:1604.06704 [hep-th]} \BibitemShut {NoStop}%
\bibitem [{\citenamefont {Cai}\ \emph {et~al.}(2022)\citenamefont {Cai}, \citenamefont {Cacciapaglia},\ and\ \citenamefont {Lee}}]{PhysRevLett.128.081806}%
  \BibitemOpen
  \bibfield  {author} {\bibinfo {author} {\bibfnamefont {H.}~\bibnamefont {Cai}}, \bibinfo {author} {\bibfnamefont {G.}~\bibnamefont {Cacciapaglia}},\ and\ \bibinfo {author} {\bibfnamefont {S.~J.}\ \bibnamefont {Lee}},\ }\bibfield  {title} {\bibinfo {title} {Massive gravitons as feebly interacting dark matter candidates},\ }\href {https://doi.org/10.1103/PhysRevLett.128.081806} {\bibfield  {journal} {\bibinfo  {journal} {Phys. Rev. Lett.}\ }\textbf {\bibinfo {volume} {128}},\ \bibinfo {pages} {081806} (\bibinfo {year} {2022})}\BibitemShut {NoStop}%
\bibitem [{\citenamefont {Rivera}(2016)}]{rivera2016}%
  \BibitemOpen
  \bibfield  {author} {\bibinfo {author} {\bibfnamefont {A.~B.}\ \bibnamefont {Rivera}},\ }\href {https://arxiv.org/abs/1601.03789} {\bibinfo {title} {Theoretical foundations of pgws printed in the cmb and its observational status}} (\bibinfo {year} {2016}),\ \Eprint {https://arxiv.org/abs/1601.03789} {arXiv:1601.03789 [astro-ph.CO]} \BibitemShut {NoStop}%
\bibitem [{202(2020)}]{2020_inflation}%
  \BibitemOpen
  \bibfield  {title} {\bibinfo {title} {Planck2018 results: X. constraints on inflation},\ }\href {https://doi.org/10.1051/0004-6361/201833887} {\bibfield  {journal} {\bibinfo  {journal} {Astron. Astrophys.}\ }\textbf {\bibinfo {volume} {641}},\ \bibinfo {pages} {A10} (\bibinfo {year} {2020})}\BibitemShut {NoStop}%
\bibitem [{\citenamefont {Tristram}\ \emph {et~al.}(2022)\citenamefont {Tristram} \emph {et~al.}}]{Tristram:2021tvh}%
  \BibitemOpen
  \bibfield  {author} {\bibinfo {author} {\bibfnamefont {M.}~\bibnamefont {Tristram}} \emph {et~al.},\ }\bibfield  {title} {\bibinfo {title} {{Improved limits on the tensor-to-scalar ratio using BICEP and Planck data}},\ }\href {https://doi.org/10.1103/PhysRevD.105.083524} {\bibfield  {journal} {\bibinfo  {journal} {Phys. Rev. D}\ }\textbf {\bibinfo {volume} {105}},\ \bibinfo {pages} {083524} (\bibinfo {year} {2022})},\ \Eprint {https://arxiv.org/abs/2112.07961} {arXiv:2112.07961 [astro-ph.CO]} \BibitemShut {NoStop}%
\bibitem [{\citenamefont {Janssen}\ \emph {et~al.}(2014)\citenamefont {Janssen}, \citenamefont {Hobbs}, \citenamefont {McLaughlin}, \citenamefont {Bassa}, \citenamefont {Deller}, \citenamefont {Kramer}, \citenamefont {Lee}, \citenamefont {Mingarelli}, \citenamefont {Rosado}, \citenamefont {Sanidas}, \citenamefont {Sesana}, \citenamefont {Shao}, \citenamefont {Stairs}, \citenamefont {Stappers},\ and\ \citenamefont {Verbiest}}]{janssen2014}%
  \BibitemOpen
  \bibfield  {author} {\bibinfo {author} {\bibfnamefont {G.~H.}\ \bibnamefont {Janssen}}, \bibinfo {author} {\bibfnamefont {G.}~\bibnamefont {Hobbs}}, \bibinfo {author} {\bibfnamefont {M.}~\bibnamefont {McLaughlin}}, \bibinfo {author} {\bibfnamefont {C.~G.}\ \bibnamefont {Bassa}}, \bibinfo {author} {\bibfnamefont {A.~T.}\ \bibnamefont {Deller}}, \bibinfo {author} {\bibfnamefont {M.}~\bibnamefont {Kramer}}, \bibinfo {author} {\bibfnamefont {K.~J.}\ \bibnamefont {Lee}}, \bibinfo {author} {\bibfnamefont {C.~M.~F.}\ \bibnamefont {Mingarelli}}, \bibinfo {author} {\bibfnamefont {P.~A.}\ \bibnamefont {Rosado}}, \bibinfo {author} {\bibfnamefont {S.}~\bibnamefont {Sanidas}}, \bibinfo {author} {\bibfnamefont {A.}~\bibnamefont {Sesana}}, \bibinfo {author} {\bibfnamefont {L.}~\bibnamefont {Shao}}, \bibinfo {author} {\bibfnamefont {I.~H.}\ \bibnamefont {Stairs}}, \bibinfo {author} {\bibfnamefont {B.~W.}\ \bibnamefont {Stappers}},\ and\ \bibinfo {author} {\bibfnamefont {J.~P.~W.}\ \bibnamefont {Verbiest}},\ }\href
  {https://arxiv.org/abs/1501.00127} {\bibinfo {title} {Gravitational wave astronomy with the ska}} (\bibinfo {year} {2014}),\ \Eprint {https://arxiv.org/abs/1501.00127} {arXiv:1501.00127 [astro-ph.IM]} \BibitemShut {NoStop}%
\bibitem [{\citenamefont {Lin}\ and\ \citenamefont {Ishak}(2016)}]{Lin:2016gve}%
  \BibitemOpen
  \bibfield  {author} {\bibinfo {author} {\bibfnamefont {W.}~\bibnamefont {Lin}}\ and\ \bibinfo {author} {\bibfnamefont {M.}~\bibnamefont {Ishak}},\ }\bibfield  {title} {\bibinfo {title} {{Testing gravity theories using tensor perturbations}},\ }\href {https://doi.org/10.1103/PhysRevD.94.123011} {\bibfield  {journal} {\bibinfo  {journal} {Phys. Rev. D}\ }\textbf {\bibinfo {volume} {94}},\ \bibinfo {pages} {123011} (\bibinfo {year} {2016})},\ \Eprint {https://arxiv.org/abs/1605.03504} {arXiv:1605.03504 [astro-ph.CO]} \BibitemShut {NoStop}%
\bibitem [{\citenamefont {Bartolo}\ \emph {et~al.}(2016)\citenamefont {Bartolo}, \citenamefont {Caprini}, \citenamefont {Domcke}, \citenamefont {Figueroa}, \citenamefont {Garcia-Bellido}, \citenamefont {Guzzetti}, \citenamefont {Liguori}, \citenamefont {Matarrese}, \citenamefont {Peloso}, \citenamefont {Petiteau}, \citenamefont {Ricciardone}, \citenamefont {Sakellariadou}, \citenamefont {Sorbo},\ and\ \citenamefont {Tasinato}}]{Bartolo_2016}%
  \BibitemOpen
  \bibfield  {author} {\bibinfo {author} {\bibfnamefont {N.}~\bibnamefont {Bartolo}}, \bibinfo {author} {\bibfnamefont {C.}~\bibnamefont {Caprini}}, \bibinfo {author} {\bibfnamefont {V.}~\bibnamefont {Domcke}}, \bibinfo {author} {\bibfnamefont {D.~G.}\ \bibnamefont {Figueroa}}, \bibinfo {author} {\bibfnamefont {J.}~\bibnamefont {Garcia-Bellido}}, \bibinfo {author} {\bibfnamefont {M.~C.}\ \bibnamefont {Guzzetti}}, \bibinfo {author} {\bibfnamefont {M.}~\bibnamefont {Liguori}}, \bibinfo {author} {\bibfnamefont {S.}~\bibnamefont {Matarrese}}, \bibinfo {author} {\bibfnamefont {M.}~\bibnamefont {Peloso}}, \bibinfo {author} {\bibfnamefont {A.}~\bibnamefont {Petiteau}}, \bibinfo {author} {\bibfnamefont {A.}~\bibnamefont {Ricciardone}}, \bibinfo {author} {\bibfnamefont {M.}~\bibnamefont {Sakellariadou}}, \bibinfo {author} {\bibfnamefont {L.}~\bibnamefont {Sorbo}},\ and\ \bibinfo {author} {\bibfnamefont {G.}~\bibnamefont {Tasinato}},\ }\bibfield  {title} {\bibinfo {title} {Science with the space-based interferometer
  lisa. iv: probing inflation with gravitational waves},\ }\href {https://doi.org/10.1088/1475-7516/2016/12/026} {\bibfield  {journal} {\bibinfo  {journal} {Journal of Cosmology and Astroparticle Physics}\ }\textbf {\bibinfo {volume} {2016}}\bibinfo  {number} { (12)},\ \bibinfo {pages} {026–026}}\BibitemShut {NoStop}%
\bibitem [{\citenamefont {Giovannini}(2014)}]{Giovannini:2014jca}%
  \BibitemOpen
\bibfield  {number} {  }\bibfield  {author} {\bibinfo {author} {\bibfnamefont {M.}~\bibnamefont {Giovannini}},\ }\bibfield  {title} {\bibinfo {title} {{Cosmic backgrounds of relic gravitons and their absolute normalization}},\ }\href {https://doi.org/10.1088/0264-9381/31/22/225002} {\bibfield  {journal} {\bibinfo  {journal} {Class. Quant. Grav.}\ }\textbf {\bibinfo {volume} {31}},\ \bibinfo {pages} {225002} (\bibinfo {year} {2014})},\ \Eprint {https://arxiv.org/abs/1405.6301} {arXiv:1405.6301 [astro-ph.CO]} \BibitemShut {NoStop}%
\bibitem [{\citenamefont {Giovannini}(2020)}]{Giovannini_2020}%
  \BibitemOpen
  \bibfield  {author} {\bibinfo {author} {\bibfnamefont {M.}~\bibnamefont {Giovannini}},\ }\bibfield  {title} {\bibinfo {title} {Primordial backgrounds of relic gravitons},\ }\href {https://doi.org/10.1016/j.ppnp.2020.103774} {\bibfield  {journal} {\bibinfo  {journal} {Progress in Particle and Nuclear Physics}\ }\textbf {\bibinfo {volume} {112}},\ \bibinfo {pages} {103774} (\bibinfo {year} {2020})}\BibitemShut {NoStop}%
\bibitem [{\citenamefont {Vagnozzi}\ and\ \citenamefont {Loeb}(2022)}]{Vagnozzi_2022}%
  \BibitemOpen
  \bibfield  {author} {\bibinfo {author} {\bibfnamefont {S.}~\bibnamefont {Vagnozzi}}\ and\ \bibinfo {author} {\bibfnamefont {A.}~\bibnamefont {Loeb}},\ }\bibfield  {title} {\bibinfo {title} {The challenge of ruling out inflation via the primordial graviton background},\ }\href {https://doi.org/10.3847/2041-8213/ac9b0e} {\bibfield  {journal} {\bibinfo  {journal} {The Astrophysical Journal Letters}\ }\textbf {\bibinfo {volume} {939}},\ \bibinfo {pages} {L22} (\bibinfo {year} {2022})}\BibitemShut {NoStop}%
\bibitem [{Note2()}]{Note2}%
  \BibitemOpen
  \bibinfo {note} {Ultra-High Frequency Gravitational Waves / Stephen Hawking Centre for Theoretical Cosmology: \protect \url {https://www.ctc.cam.ac.uk/activities/UHF-GW.php}}\BibitemShut {NoStop}%
\bibitem [{Note3()}]{Note3}%
  \BibitemOpen
  \bibinfo {note} {\protect \url {https://advancedligo.mit.edu/}}\BibitemShut {NoStop}%
\bibitem [{Note4()}]{Note4}%
  \BibitemOpen
  \bibinfo {note} {\protect \url {https://www.et-gw.eu/}}\BibitemShut {NoStop}%
\bibitem [{Note5()}]{Note5}%
  \BibitemOpen
  \bibinfo {note} {\protect \url {https://cosmicexplorer.org/}}\BibitemShut {NoStop}%
\bibitem [{Note6()}]{Note6}%
  \BibitemOpen
  \bibinfo {note} {\protect \url {https://decigo.jp/index_E.html}}\BibitemShut {NoStop}%
\bibitem [{\citenamefont {Cooray}\ and\ \citenamefont {Seto}(2004)}]{Cooray:2003cv}%
  \BibitemOpen
  \bibfield  {author} {\bibinfo {author} {\bibfnamefont {A.}~\bibnamefont {Cooray}}\ and\ \bibinfo {author} {\bibfnamefont {N.}~\bibnamefont {Seto}},\ }\bibfield  {title} {\bibinfo {title} {{Graviton mass from close white dwarf binaries detectable with LISA}},\ }\href {https://doi.org/10.1103/PhysRevD.69.103502} {\bibfield  {journal} {\bibinfo  {journal} {Phys. Rev. D}\ }\textbf {\bibinfo {volume} {69}},\ \bibinfo {pages} {103502} (\bibinfo {year} {2004})},\ \Eprint {https://arxiv.org/abs/astro-ph/0311054} {arXiv:astro-ph/0311054} \BibitemShut {NoStop}%
\bibitem [{\citenamefont {Zhao}\ \emph {et~al.}(2009)\citenamefont {Zhao}, \citenamefont {Baskaran},\ and\ \citenamefont {Coles}}]{Zhao:2009pt}%
  \BibitemOpen
  \bibfield  {author} {\bibinfo {author} {\bibfnamefont {W.}~\bibnamefont {Zhao}}, \bibinfo {author} {\bibfnamefont {D.}~\bibnamefont {Baskaran}},\ and\ \bibinfo {author} {\bibfnamefont {P.}~\bibnamefont {Coles}},\ }\bibfield  {title} {\bibinfo {title} {{Detecting relics of a thermal gravitational wave background in the early Universe}},\ }\href {https://doi.org/10.1016/j.physletb.2009.09.018} {\bibfield  {journal} {\bibinfo  {journal} {Phys. Lett. B}\ }\textbf {\bibinfo {volume} {680}},\ \bibinfo {pages} {411} (\bibinfo {year} {2009})},\ \Eprint {https://arxiv.org/abs/0907.4303} {arXiv:0907.4303 [gr-qc]} \BibitemShut {NoStop}%
\bibitem [{Note7()}]{Note7}%
  \BibitemOpen
  \bibinfo {note} {\protect \url {https://nanograv.org/}}\BibitemShut {NoStop}%
\bibitem [{Note8()}]{Note8}%
  \BibitemOpen
  \bibinfo {note} {\protect \url {https://www.epta.eu.org/}}\BibitemShut {NoStop}%
\bibitem [{Note9()}]{Note9}%
  \BibitemOpen
  \bibinfo {note} {\protect \url {https://ipta4gw.org/}}\BibitemShut {NoStop}%
\bibitem [{\citenamefont {Fang-Yu}\ and\ \citenamefont {Nan}(2004)}]{Fang_Yu_2004}%
  \BibitemOpen
  \bibfield  {author} {\bibinfo {author} {\bibfnamefont {L.}~\bibnamefont {Fang-Yu}}\ and\ \bibinfo {author} {\bibfnamefont {Y.}~\bibnamefont {Nan}},\ }\bibfield  {title} {\bibinfo {title} {Resonant interaction between a weak gravitational wave and a microwave beam in the double polarized states through a static magnetic field},\ }\href {https://doi.org/10.1088/0256-307x/21/11/011} {\bibfield  {journal} {\bibinfo  {journal} {Chinese Physics Letters}\ }\textbf {\bibinfo {volume} {21}},\ \bibinfo {pages} {2113–2116} (\bibinfo {year} {2004})}\BibitemShut {NoStop}%
\bibitem [{\citenamefont {{Baker}}\ \emph {et~al.}(2008)\citenamefont {{Baker}}, \citenamefont {{Stephenson}},\ and\ \citenamefont {{Li}}}]{2008AIPC..969.1045B}%
  \BibitemOpen
  \bibfield  {author} {\bibinfo {author} {\bibfnamefont {R.~M.~L.}\ \bibnamefont {{Baker}}}, \bibinfo {author} {\bibfnamefont {G.~V.}\ \bibnamefont {{Stephenson}}},\ and\ \bibinfo {author} {\bibfnamefont {F.}~\bibnamefont {{Li}}},\ }\bibfield  {title} {\bibinfo {title} {{Proposed Ultra-High Sensitivity High-Frequency Gravitational Wave Detector}},\ }in\ \href {https://doi.org/10.1063/1.2844941} {\emph {\bibinfo {booktitle} {Space Technology and Applications International Forum-STAIF 2008}}},\ \bibinfo {series} {American Institute of Physics Conference Series}, Vol.\ \bibinfo {volume} {969},\ \bibinfo {editor} {edited by\ \bibinfo {editor} {\bibfnamefont {M.~S.}\ \bibnamefont {{El-Genk}}}}\ (\bibinfo  {publisher} {AIP},\ \bibinfo {year} {2008})\ pp.\ \bibinfo {pages} {1045--1054}\BibitemShut {NoStop}%
\bibitem [{\citenamefont {Caprini}\ and\ \citenamefont {Figueroa}(2018)}]{Caprini_2018}%
  \BibitemOpen
  \bibfield  {author} {\bibinfo {author} {\bibfnamefont {C.}~\bibnamefont {Caprini}}\ and\ \bibinfo {author} {\bibfnamefont {D.~G.}\ \bibnamefont {Figueroa}},\ }\bibfield  {title} {\bibinfo {title} {Cosmological backgrounds of gravitational waves},\ }\href {https://doi.org/10.1088/1361-6382/aac608} {\bibfield  {journal} {\bibinfo  {journal} {Classical and Quantum Gravity}\ }\textbf {\bibinfo {volume} {35}},\ \bibinfo {pages} {163001} (\bibinfo {year} {2018})}\BibitemShut {NoStop}%
\bibitem [{\citenamefont {Chung}\ and\ \citenamefont {Li}(2021)}]{Chung_2021}%
  \BibitemOpen
  \bibfield  {author} {\bibinfo {author} {\bibfnamefont {A.~K.-W.}\ \bibnamefont {Chung}}\ and\ \bibinfo {author} {\bibfnamefont {T.~G.}\ \bibnamefont {Li}},\ }\bibfield  {title} {\bibinfo {title} {Lensing of gravitational waves as a novel probe of graviton mass},\ }\bibfield  {journal} {\bibinfo  {journal} {Physical Review D}\ }\textbf {\bibinfo {volume} {104}},\ \href {https://doi.org/10.1103/physrevd.104.124060} {10.1103/physrevd.104.124060} (\bibinfo {year} {2021})\BibitemShut {NoStop}%
\bibitem [{\citenamefont {Tobar}\ \emph {et~al.}(2024)\citenamefont {Tobar}, \citenamefont {Manikandan}, \citenamefont {Beitel},\ and\ \citenamefont {Pikovski}}]{Tobar_2024}%
  \BibitemOpen
  \bibfield  {author} {\bibinfo {author} {\bibfnamefont {G.}~\bibnamefont {Tobar}}, \bibinfo {author} {\bibfnamefont {S.~K.}\ \bibnamefont {Manikandan}}, \bibinfo {author} {\bibfnamefont {T.}~\bibnamefont {Beitel}},\ and\ \bibinfo {author} {\bibfnamefont {I.}~\bibnamefont {Pikovski}},\ }\bibfield  {title} {\bibinfo {title} {Detecting single gravitons with quantum sensing},\ }\bibfield  {journal} {\bibinfo  {journal} {Nature Communications}\ }\textbf {\bibinfo {volume} {15}},\ \href {https://doi.org/10.1038/s41467-024-51420-8} {10.1038/s41467-024-51420-8} (\bibinfo {year} {2024})\BibitemShut {NoStop}%
\end{thebibliography}%

\end{document}